\documentclass[aps,nofootinbib,showpacs,twocolumn,longbibliography]{revtex4-1}

\usepackage[CJKbookmarks, pdftex, bookmarksnumbered, bookmarksopen, colorlinks, citecolor=blue, linkcolor=blue]{hyperref}
\usepackage{amsthm}
\usepackage{graphicx}						
\usepackage{amsmath,amssymb,array}
\usepackage{amscd}
\usepackage{mathrsfs}
\usepackage{dsfont}
\usepackage{amsthm}
\usepackage{mathrsfs}
\usepackage[english]{babel}
\usepackage{enumitem} 
\usepackage{setspace}
\usepackage[]{latexsym}
\usepackage{subfigure}
\usepackage{yfonts}
\usepackage[T1]{fontenc}
\usepackage{braket}
\usepackage{float}
\usepackage{wrapfig}
\def\beqa{\begin{eqnarray}}
\def\eeqa{\end{eqnarray}}
\def\bean{\begin{eqnarray*}}
\def\eean{\end{eqnarray*}}

\newenvironment{varpmatrix}[1][\small] 
  {\left(\mbox\bgroup#1$\begin{matrix}} 
  {\end{matrix}$\egroup\right)}

\theoremstyle{plain} 

\input xy
\xyoption{all}

\begin{document}

\title{\textbf{Fisher Metric, Geometric Entanglement and Spin Networks}}

\author{Goffredo Chirco$^{1}$, Fabio M. Mele$^{2,3}$, Daniele Oriti$^{1}$ and Patrizia Vitale$^{2,3}$}
\affiliation{$^{1}$Max Planck Institute for Gravitational Physics, Albert Einstein Institute, Am M\"{u}hlenberg 1, 14476, Potsdam, Germany\\
$^{2}$ Dipartimento di Fisica, Universit\`{a} di Napoli Federico II\\
Monte S. Angelo, Via Cintia, 80126 Napoli, Italy\\
$^{3}$INFN, Sezione di Napoli, Monte S. Angelo, Via Cintia, 80126 Napoli, Italy}

%\date{\today}

\begin{abstract}
Starting from recent results on the geometric formulation of quantum mechanics, we propose a new \emph{information geometric} characterization of entanglement for spin network states in the context of quantum gravity. For the simple case of a single-link fixed graph (\emph{Wilson line}), we detail the construction of a Riemannian Fisher metric tensor and a symplectic structure on the graph Hilbert space, showing how these encode the whole information about separability and entanglement. In particular, the Fisher metric defines an \emph{entanglement monotone} which provides a notion of \emph{distance} among states in the Hilbert space. In the maximally entangled gauge-invariant case, the entanglement monotone is proportional to a power of the area of the surface dual to the link thus supporting a connection between entanglement and the (simplicial) geometric properties of spin network states. We further extend such analysis to the study of non-local correlations between two non-adjacent regions of a generic spin network graph characterized by the bipartite unfolding of an \emph{Intertwiner state}. Our analysis confirms the interpretation of spin network bonds as a result of entanglement and to regard the same spin network graph as an information graph, whose connectivity encodes, both at the local and non-local level, the quantum correlations among its parts. This gives a further connection between entanglement and geometry.

\end{abstract}

\maketitle
%\tableofcontents

\section{Introduction}

Background-independent candidates to a full theory of Quantum Gravity, such as Loop Quantum Gravity \cite{1,2,3,4,5}, the modern incarnation of the canonical quantization programme for the gravitational field, together with its covariant counterpart (spin foam models),based on simplicial gravity techniques, and Group Field Theory (GFT) \cite{6,7,8,9}, a closely related formalism sharing the same type of fundamental degrees of freedom, propose a picture of the microscopic quantum structure of spacetime, where at very small scales continuum space, time and geometry dissolve into non-geometric, combinatorial and algebraic (group-theoretic) entities. These entities can be described in terms of spin networks,  graphs coloured by irreducible representations of the local gauge group of gravity (the Lorentz group, then usually gauge-fixed to $SU(2)$). Quantum spin network states represent elementary excitations of spacetime itself, and geometric observables are operators acting on them. For example, areas and volumes correspond to quantum operators which are diagonalized on spin network states. The heuristic picture for spin networks is therefore that of ``grains of space''.

The key issue, then, becomes the reconstruction or ``emergence'' of continuum spacetime and geometry from such microscopic building blocks, in some approximate regime of the quantum dynamics. This issue is intertwined with, but goes also much beyond, the difficulties with defining a notion of locality in spacetime due to diffeomorphism invariance.

Various strategies for addressing this open issue are being explored. At a more formal level, they all aim at a better control over the regime of the fundamental theory involving a large number of fundamental degrees of freedom, and therefore rest on the renormalization of quantum gravity models. This line of research has witnessed a tremendous progress in the context of renormalization of group field theory models \cite{GFTrenorm}, which amounts automatically to a renormalization of the corresponding spin foam amplitudes, as well as in the context of spin foam models understood as generalised lattice gauge theories \cite{BiancaRenorm}. At a more physical level, the effective continuum dynamics emerging from quantum gravity models has been studied, for example, in the formalism of group field theory condensate cosmology in \cite{GFTcosmo,GFTcosmo2}, so far mostly limited to spatially homogenous and isotropic universes\footnote{A first investigation of anisotropic GFT condensate configurations can be found in \cite{GFTcosmo3} and a first step towards the analysis of inhomogeneities is taken in \cite{GFTcosmo4}.}.

There are many hints that entanglement and tools from quantum information theory should play a crucial role both in the characterization of the intrinsic properties of the quantum texture of spacetime and in the reconstruction of its geometry. For instance, in a different, but still quantum gravity-related context, recent developments in AdS/CFT have shown that the entanglement of spatial regions on the boundary is directly related to the connectivity of the bulk regions thus suggesting that our three-dimensional space is held together by quantum entanglement \cite{F4}. Later works based on the so-called \textit{Ryu-Takayanagi formula}, which relates the entanglement entropy in a conformal field theory to the area of a minimal surface in its holographic dual \cite{intr4}, have also shown that the stress-energy tensor near the boundary of a bulk spacetime region can be reconstructed from the entanglement on the boundary \cite{intr5, intr6}.

Also on the side of fundamental quantum gravity formalisms based on spin networks, there are many proposals to use quantum information to reconstruct geometrical notions such as distance in terms of the entanglement on spin network states \cite{intr7,F3,F17}. The idea is that in a purely relational, background independent context only correlations have a physical meaning and it seems reasonable to regard spin networks themselves as networks of quantum correlations between regions of space and then derive geometrical properties from the intrinsic information content of the theory. In fact, there has been also a lot of activity, recently, in connecting spin network states and tensor networks, which are a crucial tool for controlling the entanglement structure of many-body quantum states \cite{TN}, and in using the same connection to extract information about the entanglement entropy encoded in spin network states \cite{muxin, noi}. In particular, in \cite{noi}, a precise dictionary between tensor networks, spin networks and group field theory states has been established, and used as a basis for a derivation of the Ryu-Takayanagi formula in a quantum gravity context. Other results on entanglement in spin network states can be found in \cite{eugenio, horizon}. Moreover, part of the cited work on spin foam renormalization \cite{BiancaRenorm} relies as well on tensor network techniques, providing another route for understanding how the entanglement of spin network states affects the continuum limit of quantum gravity models.

This work aims at providing further insights on the transition ``\textit{from pregeometry to geometry}'' by introducing the tools of the geometric formulation of quantum mechanics (GQM) in the quantum gravity context. Indeed, the usual Hilbert space of a quantum mechanical system can be equipped with a K\"ahler manifold structure inheriting both a Riemannian metric tensor and a symplectic structure from the underlying complex projective space of rays. This is the so-called Fubini-Study Hermitian tensor whose real and imaginary parts provide us with a metric and a symplectic structure, respectively. For pure states, this metric is the Fisher-Rao metric, well known in the context of information geometry. Powerful techniques of differential geometry can be thus imported in QM and, in particular, the entanglement properties of a composite system can be characterized in a purely tensorial fashion \cite{GQM3,GQM7}. Indeed, the pulled-back Hermitian tensor on orbit submanifolds of quantum states, related by unitary transformations, decomposes in block matrices which encode all the information about the separability or entangled nature of the fiducial state of the given orbit. Of particular interest are the off-diagonal blocks of the metric which encode the information on quantum correlations between the subsystems and define an entanglement measure interpreted as a distance with respect to the separable case.

Along the same line, we can use quantum tensors to characterize the entanglement on spin network states. The advantages of this formalism are both computational and conceptual. Indeed, unlike the calculations involving entanglement entropy, it does not require the explicit knowledge of the Schmidt coefficients. Moreover, the key structures of the formalism are built purely from the space of states without introducing additional external structures. We will give further motivations for the use of such geometric techniques in a quantum gravity context, in the coming sections.\\

The paper is organized as follows. Section \ref{GQM} introduces the basics of Geometric Quantum Mechanics focusing on the characterization of entanglement by means of the tensorial structures defined on the manifold of pure quantum states. In Sections \ref{SN}-\ref{is} we try then such GQM formalism on the structure of correlations of the spin networks states, for two different simple examples, in the generalised context of GFT abstract graphs. Specifically, we first consider the case of a single-link graph regarded as a bipartite system correlating the spin states at its endpoints and explicitly discuss the two extreme cases of a separable and a maximally entangled state (Sec. \ref{wl}). Then in Section \ref{2wl} we discuss the entanglement resulting from the gluing of two links. Finally, in Sec. \ref{is} we extend the analysis to the study of non-local correlations characterizing the bipartite unfolding of an \emph{Intertwiner state}. Detailed computations and an explicit example are reported in Appendix \ref{appendixA} and \ref{appendixB}, respectively. In Section \ref{out} we collect our results and discuss future investigations.

To keep the treatment self-contained we added some further appendices at the end of the paper which give a review of the notion of spin networks in QG. We first introduce the \emph{geometric}, or \emph{embedded}, definition of states of quantum geometry as given in the canonical framework of LQG (Appendix \ref{spin}), hence we move to a generalised description provided by GFT, where graphs become abstract networks comprised by \emph{pre-geometric} quanta of space of purely combinatorial and algebraic nature (Appendix \ref{gft}).% which, because of diffeomorphism invariance, offer a realization of the pregeometric scenario in terms of building blocks of combinatorial and algebraic nature. A notion of entanglement on spin network states can be given in terms of the local correlations coming from gluing links (or surfaces in the dual picture) of open spin network vertices. In particular, the gauge-invariance requirement at the gluing nodes implies a locally maximally entangled state \cite{F1}.
%The analysis of the tensorial structures allows us to reinterpret the spin network graph structure itself as an information graph whose connectivity is encoded, both at the local and non-local level, in the choice of the entangled state (Sec. VI-VII). The entanglement measure involving the off-diagonal blocks of the metric tensor can be therefore interpreted ultimately as a measure of graph connectivity. Starting from pre-geometric spin network structures we are able to define a connection between entanglement and geometry by exploiting the geometric information carried by the states.
%}

\section{A geometric approach to Quantum Mechanics}\label{GQM}
%One of the most appealing reasons to construct a geometric formulation of Quantum Mechanics (QM) concerns the opportunity of making available ``classical methods'' of Riemannian and symplectic geometry in a quantum mechanical framework \cite{GQM1}. 
%The geometrization program for quantum mechanics can be synthesized as the replacement of the usual description of a quantum system in terms of Hilbert spaces with a description in terms of \emph{Hilbert manifolds} \cite{GQM2,GQM4,GQM6,GQM15}, allowing for the use of ``classical methods'' of Riemannian and symplectic geometry in the quantum mechanical framework \cite{GQM1}. 

According to the probabilistic interpretation of quantum mechanics, we usually identify (pure) states  with equivalence classes (\textit{rays}) of state vectors $\ket\psi$ with respect to multiplication by a non-zero complex number \cite{GQM12}. The space of rays $\mathcal R(\mathcal H)$ is a differential manifold identified with the complex projective space $\mathbb CP(\mathcal H)$ associated with $\mathcal H$ \cite{Bengts}. The manifold structure of this space requires that we replace all objects, whose definition depends on the linear structure on $\mathcal H$, with tensorial geometrical entities which preserve their meaning under general transformations and not just linear ones \cite{GQM6}.

In this section we briefly recall how to construct tensorial quantities on the space of states of a quantum system by focusing on those tensors on $\mathcal H$ which can be identified with the pull-back of tensorial objects defined on the underlying complex projective space. We therefore describe the procedure of pull-back on orbit submanifolds of quantum states with respect to the action of unitary representations of Lie groups. Eventually, we shall apply such a procedure to derive a tensorial characterization of quantum entanglement for composite systems \cite{GQM7}.  

\subsection{Classical tensors on pure states}\label{GQMa}

Let $\mathcal H\cong\mathbb C^N$ be a finite-dimensional Hilbert space of dimension $N$ and denote by $\{\ket{e_j}\}_{j=1,\dots,N}$ its (orthonormal) basis. We can introduce complex coordinate functions $\{c_j\}$ on $\mathcal H$ by setting $\braket{e_j|\psi}=c_j(\psi)$, for any $\ket\psi\in\mathcal H$. By replacing functions with their exterior differentials
we may associate with the Hermitian inner product on quantum state vectors $\braket{\cdot|\cdot}:\mathcal H\times\mathcal H\rightarrow\mathbb C$ a Hermitian covariant tensor on quantum-state-valued sections of the tangent bundle $T\mathcal H$ defined by \cite{GQM2,GQM4,GQM15}
\begin{equation}\label{hermitian}
h=\braket{d\psi\otimes d\psi}:=\sum_jd\bar{c}_j\otimes d c_j\,,
\end{equation}
such that
\begin{equation}
\braket{d\psi\otimes d\psi}(X_\psi,X_{\psi'})=\braket{\psi|\psi'}
\end{equation}
for any vector field $X_\psi:\phi\mapsto(\phi,\psi),\;\forall\phi\in\mathcal H$. This essentially amounts to identify $\mathcal H$ with the tangent space $T_\phi\mathcal H$ at each point of the base manifold.

The decomposition of the coordinate functions $c_j$ into real and imaginary part, say $c_j=x_j+iy_j$, that is to replace $\mathcal H$ with its realification $\mathcal H_\mathbb R:=\mathbb Re(\mathcal H)\oplus\mathbb Im(\mathcal H)\cong\mathbb R^{2N}$, allows to identify an Euclidean metric and a symplectic structure on $\mathcal H_\mathbb R$ respectively with the real and imaginary part of the Hermitian tensor (\ref{hermitian}), i.e.
\begin{align}
h=g+i\omega=&\delta_{jk}(dx^j\otimes dx^k+dy^j\otimes dy^k)+\\ \nonumber
&\qquad +i\delta_{jk}(dx^j\otimes dy^k-dy^j\otimes dx^k)\;.
\end{align}
These two tensors are related by a $(1,1)$-tensor field $J=\delta_{jk}\bigl(dx^j\otimes\frac{\partial}{\partial y^k}-dy^j\otimes\frac{\partial}{\partial x^k}\bigr)$ playing the role of a complex structure. The real differential manifold $\mathcal H_\mathbb R$ is thus equipped with a K\"ahler manifold structure \cite{GQM15}.

Coming back to the space of rays, it is well known \cite{Bengts,GQM12} that the equivalence classes of state vectors identifying points of the complex projective space $\mathbb CP(\mathcal H)\cong\mathcal R(\mathcal H)$ can be represented by rank-one projectors $\rho=\frac{\ket\psi\bra\psi}{\braket{\psi|\psi}}\in D^1(\mathcal H)\subset\mathfrak{u}^*(\mathcal H)$ called \textit{pure states} which satisfy the properties $\rho^\dagger=\rho, \rho^2=\rho, \text{Tr}\rho=1$. Inheriting the differential calculus from $\mathfrak{u}^*(\mathcal H)$, we define an operator-valued (0,2)-tensor $d\rho\otimes d\rho$ which may be turned into a covariant tensor by evaluating it on the state $\rho$ itself, i.e.
\begin{equation}\label{FS}
\text{Tr}\left(\rho d\rho\otimes d\rho\right)\;.
\end{equation}
The pull-back of the tensor (\ref{FS}) from $\mathcal R(\mathcal H)$ to  $\mathcal H_0\equiv\mathcal H-\{\mathbf0\}$ along the (momentum) map
\begin{equation}
\mu\;:\;\mathcal H_0\ni\ket\psi\longmapsto\rho=\frac{\ket\psi\bra\psi}{\braket{\psi|\psi}}\in\mathcal R(\mathcal H)\cong D^1(\mathcal H)\subset\mathfrak{u}^*(\mathcal H) \label{f3}
\end{equation}
gives the so-called Fubini-Study Hermitian tensor \cite{GQM16}
\begin{equation}\label{fstensor}
h_{FS}=\frac{\braket{d\psi \otimes d\psi}}{\braket{\psi|\psi}}-\frac{\braket{\psi|d\psi}}{\braket{\psi|\psi}}\otimes\frac{\braket{d\psi|\psi}}{\braket{\psi|\psi}}\;,
\end{equation}
whose real and imaginary parts define a metric and a symplectic structure on $\mathcal H_0$. Therefore, according to the diagram
\begin{equation}\label{gqm44}
\begin{CD}
\mathcal{H}_0@>\mu>>\mathfrak{u}^*(\mathcal{H})\\
@V\pi VV               @AAi A\\
\mathcal{R}(\mathcal H)@>>\cong>D^1(\mathcal H)
\end{CD}
\end{equation}
the space of pure quantum states naturally inherits a K\"ahler structure from $\mathcal H_0$.

In particular, one can exploit the above construction to describe specific manifolds of states of the quantum system under consideration. Examples of manifolds of quantum states are provided for instance by coherent states \cite{gencohe} or by the stratified manifold of density states, where each stratus contains density states with fixed rank \cite{GQM6}. Indeed, given a finite-dimensional\footnote{Due to the present state-of-the art of infinite dimensional differential geometry, methods from differential geometry are much more effective when the identified submanifold has finite dimension. Fortunately many situations of great physical interest like those emerging in quantum computation are concerned with finite dimensional manifolds of quantum states.} manifold  $\mathbb M$  and $i_\mathbb M:\mathbb M\hookrightarrow\mathcal H$ the embedding of $\mathbb M$ into $\mathcal H$, the induced pull-back $i^*_\mathbb M$ of the Hermitian tensor (\ref{FS}) or (\ref{fstensor}) defines a covariant Riemannian metric tensor and a closed (symplectic in a non-degenerate case) 2-form on $\mathbb M$.

In this spirit it has been shown in \cite{GQM9} that, if $\mathbb M$ is the space (of parameters) of probability distributions associated to quantum states, the Fisher-Rao metric tensor used in statistics and information theory \cite{GQM23} can be obtained from the Fubini-Study tensor defined on the space of pure quantum states. In what follows we will then take (\ref{FS}) to be the definition of the quantum Fisher tensor.

A convenient way to identify submanifolds of quantum states, which will turn very useful for characterizing entanglement of composite systems, consists in considering orbits originated from some fiducial state. Specifically, if $\mathbb M$ admits the structure of a Lie group $\mathbb G$, the orbits\footnote{Here $\mathbb G_0$ is the isotropy group of the state $\ket0$, i.e. the subgroup of elements of $\mathbb G$ which leave the state $\ket0$ unchanged, and $\sim$ is the equivalence relation with respect to such an action.}
\begin{equation}
\mathcal O\cong\mathbb G\bigl/\mathbb G_0=\bigl\{\ket g=U(g)\ket0\;|\;g\in\mathbb G\bigr\}\bigl/\sim\;\;,
\end{equation}
generated by the action of a unitary representation $U(g)$ of $\mathbb G$ upon a normalized fiducial state $\ket0\in\mathcal H_0$ identify submanifolds of quantum states $\ket g$ when we consider an embedding map via the group action
\begin{equation}
\phi_0:\mathbb G\ni g\longmapsto\ket g=U(g)\ket0\in\mathcal H_0\;.
\end{equation}

Correspondingly on $\mathcal R(\mathcal H)$ we identify orbit submanifolds of pure quantum states with respect to the co-adjoint action on some fiducial  pure state $\rho_0=\frac{\ket0\bra0}{\braket{0|0}}$

\begin{equation}
\tilde\phi_0:\mathbb G\ni g\longmapsto\rho(g)=U(g)\rho_0U^{-1}(g)\in\mathcal R(\mathcal H)\;.
\end{equation}

We can thus restrict ourselves to the Hermitian tensor on this submanifold by noticing that it is completely described by the pull-back tensor on the Lie group $\mathbb G$ according to the following diagrams 
\begin{equation}\label{gqm89}
\begin{CD}
\mathbb{G}@>\phi_0>>\;\mathcal H_0\;\\
@V\pi_0 VV               @AAi_\mathcal O A\\
\mathbb G\bigl/\mathbb G_0@>\cong>>\;\mathcal O
\end{CD}\quad\Longrightarrow\quad\begin{CD}
\mathbb{G}@>\phi_0>>\;S(\mathcal H)\;\\
@VU(1) VV               @VVU(1) V\\
\mathbb{G}/U(1)@>\tilde\phi_0>>\;\mathcal R(\mathcal H)\;\\
@V\pi_0 VV               @AAi_\mathcal O A\\
\mathbb G\bigl/\mathbb G_0^{U(1)}@>\cong>>\;\mathcal O_{\rho_0}
\end{CD}
\end{equation}

\noindent
\\where $S(\mathcal H)=\{\bigl|\psi\bigr>\in\mathcal H_0:\bigl<\psi\big|\psi\bigr>=1\}\subset\mathcal H_0$ is the unit sphere of normalized state vectors and $\mathbb G_0^{U(1)}$ is an enlarged isotropy group taking into account the $U(1)$-degeneracy directions for the Hermitian tensor on pure states.\\ \\The operator-valued 1-form $d\rho$ can be thus written as
\begin{align}\label{drho}
d\rho&=dU\rho_0U^{-1}+U\rho_0dU^{-1}+Ud\rho_0U^{-1}\\ \nonumber
&=U[U^{-1}dU,\rho_0]_-U^{-1}\;,
\end{align}
$d\rho_0=0$ being $\rho_0$ our fixed fiducial state. The pull-back of the Hermitian tensor (\ref{FS}) on the orbit submanifold embedded in $D^1(\mathcal H)\cong\mathcal R(\mathcal H)$ then yields \cite{GQM4,GQM7}

\begin{align}\label{orbitensor}
\mathcal K&=\left\{\text{Tr}\bigl(\rho_0R(X_j)R(X_k)\bigr)-\right. \\ \nonumber
&\qquad - \left. \text{Tr}\bigl(\rho_0R(X_j)\bigr)\text{Tr}\bigl(\rho_0R(X_k)\bigr)\right\}\theta^j\otimes\theta^k\;,
\end{align}

where $R(X_j)$ denotes the Lie algebra representation defined by the unitary representation of the Lie group $\mathbb G$ and $\theta^j$ the dual basis of left-invariant 1-forms such that $U^{-1}dU= iR(X_j)\theta^j$.

Again, the real symmetric and imaginary skewsymmetric part of the tensor $\mathcal K$ provide a Riemannian metric tensor and a symplectic structure on the orbit $\mathcal O_{\rho_0}$, respectively given by\footnote{$[\cdot,\cdot]_{\pm}$ respectively denote the anticommutator and the commutator and we use the shorthand notation
$$
\theta^a\odot \, \theta^b=\frac{1}{2}\bigl(\theta^a\otimes\theta^b+\theta^b\otimes\theta^a\bigr),\quad\theta^a\wedge \theta^b=\frac{1}{2}\bigl(\theta^a\otimes\theta^b-\theta^b\otimes\theta^a\bigr)\;.
$$
for the symmetrized and antisymmetrized product of forms.
} :
\begin{align}
\mathcal K_+&=\left \{\frac{1}{2}\text{Tr}\bigl(\rho_0\left[R(X_j),R(X_k)\right]_+\bigr)-\right. \\ \nonumber
&\qquad - \text{Tr}\bigl(\rho_0R(X_j)\bigr)\text{Tr}\bigl(\rho_0R(X_k)\bigr) \bigg \}\theta^j\odot\theta^k\;,\\
\mathcal K_-&=\frac{1}{2}\text{Tr}\bigl(\rho_0\left[R(X_j),R(X_k)\right]_-\bigr)\theta^j\wedge\theta^k\;.
\end{align}

\subsection{Quantum Fisher tensor for bipartite N-level systems}\label{GQMb}

Such a pull-back procedure can be extended also to the case of a composite system \cite{GQM3,GQM7}. The correlation properties of the fiducial state $\rho_0$ are captured by the tensorial structures induced on the orbits of the action of local unitary groups which define submanifolds of states with fixed amount of entanglement. \\ 

Let $\mathcal H=\mathcal H_A\otimes\mathcal H_B\cong\mathbb C^{N_A}\otimes\mathbb C^{N_B}$ be the Hilbert space of a composite system consisting of two $N$ level systems $A$ and $B$ with number of levels respectively given by $N_A=\text{dim}\,\mathcal H_A$ and $N_B=\text{dim}\,\mathcal H_B$. For the sake of clarity, in what follows we will denote by $\otimes$ the usual tensor product of spaces and by $\otimes_F$ the product of forms. So let $\rho_0$ be a fiducial pure state in $D^1(\mathcal H_A\otimes\mathcal H_B)$ and $\mathbb G_0$ its isotropy group, we want to compute now the pull-back of the Hermitian quantum Fisher tensor
\begin{equation}\label{fisherr1}
\text{Tr}(\rho\,d\rho\underset{F}{\otimes}d\rho)
\end{equation}
on the orbits
\begin{equation}\label{fisherr2}
\mathcal O\cong U(N_A)\times U(N_B)\bigl/\mathbb G_0
\end{equation}
of unitarily related (pure) quantum states $\rho=U\rho_0U^{-1}$, induced by the co-adjoint action of the unitary group on $\rho_0$ with respect to the product representation
\begin{equation}\label{fisher4}
U=U_A\otimes U_B=(U_A\otimes\mathds1_B)\cdot(\mathds1_A\otimes U_B)\;.
\end{equation}
The corresponding Lie algebra representation $\mathfrak u(\mathcal H_A)\oplus\,\mathfrak u(\mathcal H_B)$ is provided by means of the following realization
\begin{equation}\label{fisher5}
R(X_j)=\begin{cases}
\sigma_j^{(A)}\otimes\mathds1_B\qquad\text{for}\;\;1\leq j \leq N_A^2\\
\mathds1_A\otimes\sigma_{j-N_A^2}^{(B)}\,\quad\text{for}\;\;N_A^2+1\leq j\leq N_A^2+N_B^2
\end{cases}
\end{equation}
of the infinitesimal generators of the one-dimensional subgroup of $U(N_A)\times U(N_B)$. In the following, for both subsystems we will adopt a short-hand notation of indices $a,b$ without specifying their range of values.  Therefore, being $U^{-1}dU$ a left-invariant 1-form, it can be decomposed as\footnote{Here we use the decomposition of the exterior differential operator $d=d_A\otimes\mathds1_B+\mathds1_A\otimes d_B$ acting on a product representation (\ref{fisher4}).}
\begin{equation}\label{fisher7}
U^{-1}dU=i\sigma_a^{(A)}\theta_A^a\otimes\mathds1_B+\mathds1_A\otimes i\sigma_b^{(B)}\theta_B^b\;,
\end{equation}
where $\{\theta_A\}$ and $\{\theta_B\}$ denote a basis of left-invariant 1-forms on the corresponding Lie group representation acting on the subsystem $A,B$ respectively. The operator-valued 1-form (\ref{drho}) can be thus written as
\begin{equation}\label{fisher8}
d\rho=U\bigl[i\sigma_a^{(A)}\theta_A^a\otimes\mathds1_B,\rho_0\bigr]_-U^{-1}+U\bigl[\mathds1_A\otimes i\sigma_b^{(B)}\theta_B^b,\rho_0\bigr]_-U^{-1}\;,
\end{equation}
from which it follows that
%\begin{align} \label{fisher9}
%&d\rho\underset{F}{\otimes}d\rho=\\ \nonumber
%&=\left(U\bigl[i\sigma_a^{(A)}\otimes\mathds1_B,\rho_0\bigr]_-\bigl[i\sigma_b^{(A)}\otimes\mathds1_B,\rho_0\bigr]_-U^{-1}\right)\theta_A^a\underset{F}{\otimes}\theta_A^{b}\nn \\
%&+\left(U\bigl[i\sigma_a^{(A)}\otimes\mathds1_B,\rho_0\bigr]_-\bigl[\mathds1_A\otimes i\sigma_b^{(B)},\rho_0\bigr]_-U^{-1}\right)\theta_A^a\underset{F}{\otimes}\theta_B^{b}\nn\\
%&+\left(U\bigl[\mathds1_A\otimes i\sigma_a^{(B)},\rho_0\bigr]_-\bigl[i\sigma_b^{(A)}\otimes\mathds1_B,\rho_0\bigr]_-U^{-1}\right)\theta_B^a\underset{F}{\otimes}\theta_A^{b}\nn\\
%&+ \left(U\bigl[\mathds1_A\otimes i\sigma_a^{(B)},\rho_0\bigr]_-\bigl[\mathds1_A\otimes i\sigma_b^{(B)},\rho_0\bigr]_-U^{-1}\right)\theta_B^a\underset{F}{\otimes}\theta_B^{b}\;.\nonumber
%\end{align}
the pull-back of the Hermitian tensor (\ref{fisherr1}) on the orbit submanifold starting from the fiducial state $\rho_0$ then reads as
\begin{align}\label{fisher10}
\mathcal K&=\mathcal K_{ab}^{(A)}\theta_A^a\underset{F}{\otimes}\, \theta_A^b+\mathcal K_{ab}^{(AB)}\theta_A^a\underset{F}{\otimes}\, \theta_B^b+\\ \nonumber
&\qquad \mathcal + \,K_{ab}^{(BA)}\theta_B^a\underset{F}{\otimes}\,\theta_A^b+\mathcal K_{ab}^{(B)}\theta_B^a\underset{F}{\otimes}\,\theta_B^b
\end{align}
with
\begin{equation}\label{fisher11}
\begin{cases}
\mathcal K_{ab}^{(A)}=-\text{Tr}\bigl(\rho_0\bigl[\sigma_a^{(A)}\otimes\mathds1_B,\rho_0\bigr]_-\bigl[\sigma_b^{(A)}\otimes\mathds1_B,\rho_0\bigr]_-\bigr)\\
\mathcal K_{ab}^{(AB)}=-\text{Tr}\bigl(\rho_0\bigl[\sigma_a^{(A)}\otimes\mathds1_B,\rho_0\bigr]_-\bigl[\mathds1_A\otimes\sigma_b^{(B)},\rho_0\bigr]_-\bigr)\\
\mathcal K_{ab}^{(BA)}=-\text{Tr}\bigl(\rho_0\bigl[\mathds1_A\otimes\sigma_a^{(B)},\rho_0\bigr]_-\bigl[\sigma_b^{(A)}\otimes\mathds1_B,\rho_0\bigr]_-\bigr)\\
\mathcal K_{ab}^{(B)}=-\text{Tr}\bigl(\rho_0\bigl[\mathds1_A\otimes\sigma_a^{(B)},\rho_0\bigr]_-\bigl[\mathds1_A\otimes\sigma_b^{(B)},\rho_0\bigr]_-\bigr)\;.
\end{cases}
\end{equation}
Now, being $\rho_0^3=\rho_0^2=\rho_0$ for a pure state $\rho_0$, a direct computation shows that 
%\begin{align}\label{fisher12}
%&\mathcal K_{ab}^{(A)}=\text{Tr}\bigl(\rho_0\sigma_a^{(A)}\sigma_b^{(A)}\otimes\mathds1_B\bigr)- \\ \nonumber
%&\qquad \qquad -\, \text{Tr}\bigl(\rho_0\sigma_a^{(A)}\otimes\mathds1_B\bigr)\text{Tr}\bigl(\rho_0\sigma_b^{(A)}\otimes\mathds1_B\bigr)\;,\\
%&\mathcal K_{ab}^{(AB)}=\text{Tr}\bigl(\rho_0\sigma_a^{(A)}\otimes\sigma_b^{(B)}\bigr)- \\ \nonumber
%&\qquad \qquad -\,\text{Tr}\bigl(\rho_0\sigma_a^{(A)}\otimes\mathds1_B\bigr)\text{Tr}\bigl(\rho_0\mathds1_A\otimes\sigma_b^{(B)}\bigr)\;,\\
%&\mathcal K_{ab}^{(BA)}=\text{Tr}\bigl(\rho_0\sigma_a^{(B)}\otimes\sigma_b^{(A)}\bigr)- \\ \nonumber
%&\qquad \qquad -\,\text{Tr}\bigl(\rho_0\sigma_a^{(B)}\otimes\mathds1_A\bigr)\text{Tr}\bigl(\rho_0\mathds1_B\otimes\sigma_b^{(A)}\bigr)\;,\\
%&\mathcal K_{ab}^{(B)}=\text{Tr}\bigl(\rho_0\mathds1_A\otimes\sigma_a^{(B)}\sigma_b^{(B)}\bigr)- \\ \nonumber
%&\qquad \qquad -\,\text{Tr}\bigl(\rho_0\mathds1_A\otimes\sigma_a^{(B)}\bigr)\text{Tr}\bigl(\rho_0\mathds1_A\otimes\sigma_b^{(B)}\bigr)\;.
%\end{align}
the pulled-back Hermitian tensor $\mathcal K$ decomposes into a Riemannian metric $\mathcal K_{+}$ and a symplectic structure $\mathcal K_-$
\begin{align}\label{fisher20}
\mathcal K&=\mathcal K_+ + i\,\mathcal K_-=\\ \nonumber&=\left(\begin{array}{c|c}
\mathcal K^{(A)}_{(ab)} & \mathcal K^{(AB)}_{(ab)} \\
\hline
\mathcal K^{(BA)}_{(ab)} & \mathcal K^{(B)}_{(ab)}
\end{array}
\right)+i\,\left(\begin{array}{c|c}
\mathcal K^{(A)}_{[ab]} & 0 \\
\hline
0 & \mathcal K^{(B)}_{[ab]}
\end{array}
\right)\;,
\end{align}
%respectively given by
%\begin{align}\label{fisher17}
%\mathcal K_+&=\mathcal K_{(ab)}^{(A)}\theta_A^a\odot\theta_A^b+\mathcal K_{(ab)}^{(AB)}\theta_A^a\odot\theta_B^b+ \\ \nonumber
%&\qquad +\,\mathcal K_{(ab)}^{(BA)}\theta_B^a\odot\theta_A^b+ \mathcal K_{(ab)}^{(B)}\theta_B^a\odot\theta_B^b\;, \\
%\mathcal K_-&=\mathcal K_{[ab]}^{(A)}\theta_A^a\wedge\theta_A^b+\mathcal K_{[ab]}^{(B)}\theta_B^a\wedge\theta_B^b\;, 
%\end{align}
with
\begin{align}\label{fisher19}
\begin{cases}
&\mathcal K_{(ab)}^{(A)}=\frac{1}{2}\text{Tr}\left(\rho_0\bigl[\sigma_a^{(A)},\sigma_b^{(A)}\bigr]_+\otimes\mathds1_B\right)- \\ 
&\qquad -\,\text{Tr}\left(\rho_0\sigma_a^{(A)}\otimes\mathds1_B\right)\text{Tr}\left(\rho_0\sigma_b^{(A)}\otimes\mathds1_B\right)\\
&\mathcal K_{(ab)}^{(B)}=\frac{1}{2}\text{Tr}\left(\rho_0\mathds1_A\otimes\bigl[\sigma_a^{(B)},\sigma_b^{(B)}\bigr]_+\right)- \\ 
&\qquad -\,\text{Tr}\left(\rho_0\mathds1_A\otimes\sigma_a^{(B)}\right)\text{Tr}\left(\rho_0\mathds1_A\otimes\sigma_b^{(B)}\right)
\end{cases}
\end{align}

\begin{align}\label{fisher19a}
\begin{cases}
&\mathcal K_{(ab)}^{(AB)}=\text{Tr}\left(\rho_0\sigma_a^{(A)}\otimes\sigma_b^{(B)}\right)- \\ 
&\qquad -\,\text{Tr}\left(\rho_0\sigma_a^{(A)}\otimes\mathds1_B\right)\text{Tr}\left(\rho_0\mathds1_A\otimes\sigma_b^{(B)}\right)\\
&\mathcal K_{(ab)}^{(BA)}=\text{Tr}\left(\rho_0\sigma_a^{(B)}\otimes\sigma_b^{(A)}\right)- \\ 
&\qquad -\,\text{Tr}\left(\rho_0\sigma_a^{(B)}\otimes\mathds1_A\right)\text{Tr}\left(\rho_0\mathds1_B\otimes\sigma_b^{(A)}\right)
\end{cases}
\end{align}
and
\begin{align}\label{fisher199}
\begin{cases}
&\mathcal K_{[ab]}^{(A)}=\frac{1}{2}\text{Tr}\left(\rho_0\bigl[\sigma_a^{(A)},\sigma_b^{(A)}\bigr]_-\otimes\mathds1_B\right)\\
&\mathcal K_{[ab]}^{(B)}=\frac{1}{2}\text{Tr}\left(\rho_0\mathds1_A\otimes\bigl[\sigma_a^{(B)},\sigma_b^{(B)}\bigr]_-\right)
\end{cases}
\end{align}
where we used the bracket notation $(ab),[ab]$ for the matrix indices of the symmetric and antisymmetric part, respectively. The coefficient matrix of the Hermitian tensor (\ref{fisher20}) splits into different blocks carrying the information about the separable or entangled nature of the fiducial state $\rho_0$.
%\begin{equation}\label{fisher20}
%\mathcal K_{ab}=\left(\begin{array}{c|c}
%\mathcal K^{(A)}_{(ab)} & \mathcal K^{(AB)}_{(ab)} \\
%\hline
%\mathcal K^{(BA)}_{(ab)} & \mathcal K^{(B)}_{(ab)}
%\end{array}
%\right)+i\,\left(\begin{array}{c|c}
%\mathcal K^{(A)}_{[ab]} & 0 \\
%\hline
%0 & \mathcal K^{(B)}_{[ab]}
%\end{array}
%\right)\;.
%\end{equation}
Indeed, by means of the Fano decomposition of $\rho_0$, it is easy to see that \cite{GQM7}
\begin{itemize}
\item when $\rho_0$ is separable the off-diagonal blocks of the metric component vanish and the pulled-back Hermitian tensor $\mathcal K$ decomposes into a direct sum $\mathcal K_A\oplus\mathcal K_B$ of Hermitian tensors associated with the two subsystems;
\item when $\rho_0$ is maximally entangled the symplectic component vanishes.
\end{itemize}
In particular, the information about (quantum) correlations between the two subsystems is encoded in the off-diagonal block-coefficient $N_A^2\times N_B^2$ and $N_B^2\times N_A^2$ matrices $\mathcal K^{(AB)}$ and $\mathcal K^{(BA)}$ which allow us to define an \emph{entanglement monotone} given by \cite{GQM26}
\begin{align}\label{fisher21}
\mathcal E&=\frac{N_<^2}{4(N_<^2-1)}\text{Tr}\left(\mathcal K^{(AB)\,T}\mathcal K^{(AB)}\right)=\\ \nonumber
&=\frac{N_<^2}{4(N_<^2-1)}\text{Tr}\left(\mathcal K^{(BA)\,T}\mathcal K^{(BA)}\right)\;,
\end{align}
where $N_<=\min(N_A, N_B)$. Such a measure of entanglement is directly related to a geometric definition of distance between entangled and separable states introduced in \cite{GQM10}:
\begin{equation}\label{fisher22}
\Sigma=\text{Tr}\left(R^\dagger R\right)\quad \text{with} \quad R:=\rho_0-\rho_0^{(A)}\otimes\rho_0^{(B)}\;,
\end{equation}
and $\rho_0^{(A,B)}=\text{Tr}_{B,A}(\rho_0)$ the reduced states. Indeed, as shown in \cite{GQM10}, the geometric distance $\Sigma$ can be computed to be
\begin{equation}\label{geomdist}
\Sigma=\frac{1}{N_<^4}\text{Tr}\left(\mathcal K^{(AB)\,T}\mathcal K^{(AB)}\right)\;,
\end{equation}
thus giving a geometrical interpretation of $\mathcal E$ itself.

%\textcolor{blue}{COMMENT: Give a first comment here on the interest of defining distances wrt density matrices, playing the role of quantum metrices. Perspective particularly well suited for background independent approaches, etc... }
 
\section{Geometry of quantum spin network states}\label{SN}

In the microscopic description of spacetime provided by the background-independent approaches to quantum gravity, \cite{1,2,3,4,5,6,7,8,9}, quantum states of space geometry are described in terms of spin networks \cite{1,2,3}. In the language of tensor networks \cite{TN}, spin networks are symmetric tensor network states given by collections of quantum tensor states characterised by $3$d rotation invariance (the Lorentz group in LQG, usually gauge-fixed to $SU(2)$) linked to each other by group holonomy actions encoding the change of frame from one tensor to the next.  This set of frame transformations translates into graphs coloured by irreducible representations of the local gauge group. In the end, this leaves with a graph with group representations associated to its links, with group intertwiners associated to its nodes, which is the usual characterization of spin network states \cite{1,2,3}. 

In the context of LQG, spin networks are by construction embedded into continuum 3d manifolds, from which they partly inherit a natural geometric characterisation. In related approaches, like Tensor Models or Group Field Theories, as well as in spin foam models based on simplicial gravity ideas, spin networks associated to emerging random geometries are not embedded, hence they must be interpreted as \emph{abstract graphs} defined with no reference to any background notions of space, time or geometry \cite{LQG60}, and can be at best associated with quantized simplicial (piecewise-flat) geometries\footnote{The possibility of defining spin network states in a more abstract, combinatorial way has been considered also within the canonical LQG approach \cite{absn1,absn2}.}. 
A short review of both the LQG embedded and GFT abstract spin networks description is given in Appendix \ref{spin}. 

In absence of a background metric structure, and due to the dynamical nature of any additional discrete \lq quantum geometric\rq variable that can be associated to the spin network graph, adjacent regions of a spin network will not necessarily correspond to regions of space that are \lq close\rq in a geometric sense. Moreover, 3d spatial geometry will generically be realised as a quantum superposition of abstract non-embedded entities each of which having a different connectivity (i.e., a different graph structure), what is local in one term of the superposition, with respect to the combinatorial and algebraic data characterizing it, will in general not be local in others \cite{LQG56,LQG57}. 
Still, a given region of a spin network can be localized in a combinatorial sense, with respect to other parts of the graph.

In this framework, the general algorithmic procedure to construct tensorial geometric structures on the space of states of a given quantum theory becomes a new crucial tool to reconsider notions as  ``close'' and ``far'' in terms of quantum correlations between subregions of the spin network graph or more generally as relations between different spin network configurations (states) in the Hilbert space. This is the main focus of our work.

Given the highly intricate structure of the spin network Hilbert space, as a preliminary step along this line, we start our analysis by focusing on a set of states which constitute the fundamental building blocks of the spin network description, and more generally of any tensor network representation of lattice gauge theory: the Wilson line states and the intertwiner states, respectively providing the basic structure for links and nodes of the network.

\section{Fubini--Study Tensor for the Single link state}\label{wl}

The correlation structure of spin network states is encoded in the connectivity of the underlying graphs, which translates into entanglement between the fundamental vertex states connected by links. 
To a given graph $\Gamma$ (see Appendix \ref{spin}) is associated a Hilbert space $\mathcal{H}_{\Gamma} \cong L^2[SU(2)^L]$, where $L$ indicates the number of links comprising the graph. 
A basis for $\mathcal{H}_\Gamma$ can be naturally derived starting from the Peter-Weyl theorem \cite{LQG31}, which gives the unitary equivalence
\begin{equation}
L^2[SU(2)^L] \cong \bigotimes_{\ell=1}^L \bigoplus_{\{\vec{j}\}} \mathcal{V}^{(j_\ell)} \otimes \mathcal{V}^{(j_\ell)*}\quad,
\end{equation}
where $\mathcal V^{(j)}$ denotes the $(2j+1)$-dimensional linear space carrying the irreducible representation of $SU(2)$ and for any $j\in\frac{\mathbb N}{2}$, the system $\{\ket{j,m}\}_{-j\leq m \leq j}$ is orthonormal, i.e.
\begin{equation}\label{f35}
\mathcal V^{(j)}=span\{\ket{j,m}\}_{-j\leq m \leq j}\;,
\end{equation}
while $\mathcal V^{(j)*}$ is its dual vector space.
A function $\psi_{\Gamma}\in\mathcal H_\Gamma$ can be decomposed as
\begin{equation}\label{qg62}
\psi_{\Gamma}=\sum_{j_\ell,m_\ell,n_\ell}f_{m_1,\dots,m_L,n_1,\dots,n_L}D^{(j_1)}_{m_1n_1}(h_1)\dots D^{(j_L)}_{m_Ln_L}(h_L)\;,
\end{equation}
where $D^{(j_\ell)}_{m_\ell n_\ell}(h)=\braket{j_\ell,m_\ell|D^{(j_\ell)}(h)|j_\ell,n_\ell}$ are the Wigner D-matrix elements corresponding to the spin-$j$ irreducible representations of the group elements $h_\ell\in SU(2)$ labelling the link. An orthonormal basis for the Hilbert space $\mathcal H_\Gamma$ is thus provided by 
\begin{equation}\label{qg63}
\braket{\vec{h}|\Gamma; \vec j, \vec m, \vec n}\equiv\left(\prod_{\ell=1}^L\sqrt{2j_\ell+1}\right)D^{(j_1)}_{m_1n_1}(h_1)\dots D^{(j_L)}_{m_Ln_L}(h_L)\;,
\end{equation}
where the compact vectorial notation $\vec j, \vec m, \vec n$ denotes the spin labels of the unitary irreducible representations of $SU(2)$ associated with each link of the graph, and similarly for the corresponding group elements. 
The simplest case of a single link graph $\gamma$ is given by the state
\begin{equation}\label{f4}
\ket{\psi_\gamma}=\sum_{jmn}c^{j}_{mn}\ket{j,m,n}\;\in\;\mathcal H_\gamma \cong L^2[SU(2)]\;,
\end{equation}
where the generic Wilson line state $|j,m,n\rangle$ defines the matrix element of the representation of the holonomy along the link,
\begin{equation}\label{f8}
\braket{h|j,m,n}:=\sqrt{2j+1}\,D^{(j)}_{mn}(h)\;.
\end{equation}
The orthogonality relations of the Wigner representation matrices $D^{(j)}_{mn}$ ensure the normalization of the basis states
\begin{equation}\label{f9}
\braket{j',m',n'|j,m,n}=\delta_{jj'}\delta_{mm'}\delta_{nn'}\;,
\end{equation}
together with the decomposition of the identity
\begin{equation}\label{f10}
\mathds1=\sum_{jmn}\ket{j,m,n}\bra{j,m,n}\;.
\end{equation}
Any state in $\mathcal H_\gamma$ can be therefore expanded in the spin basis as in (\ref{f4}) with coefficients given by
\begin{align}\label{f11}
c_{mn}^{j}\equiv\braket{j,m,n|\psi_\gamma}&=\int dh\,\psi_\gamma[h]\braket{j,m,n|h}
\\ \nonumber
&=\sqrt{2j+1}\int dh\,\psi_\gamma[h]\,\overline{D_{mn}^{(j)}(h)}\;.
\end{align}
Working in the spin basis will be convenient to express our results in terms of the algebraic data of the spin-network graph ($j, m$ and $n$ in this specific situation). 

Now, by recalling the notation adopted in Section \ref{GQM}, for the link state in spin basis we can associate 
\begin{align}\label{f13}\nonumber
&\ket{e_a}\;\;\;\quad\longleftrightarrow\quad\;\ket{j,m,n}\equiv\ket{e^{(j)}_{mn}} \\ \nonumber
%\label{f14}
&\,c_a(\psi)\;\;\quad\longleftrightarrow \quad c^{(j)}_{mn}\equiv\braket{j,m,n|\psi_\gamma}\\ \nonumber
%\label{f15}
&\ket{d\psi}=\sum_adc_a\ket{e_a}\longleftrightarrow\;\;\ket{d\psi_\gamma}=\sum_{jmn}dc^{j}_{mn}\ket{j,m,n} \nonumber
\end{align}
and thereby, derive
%\begin{align}\label{f16}
%\braket{d\psi_\gamma^{(j)}\otimes d\psi_\gamma^{(j)}}=d\overline{c}^{(j)}_{mn}\otimes dc_{mn}^{(j)}\quad(\text{sum over j,m,n})\;
%\end{align}
%given that we can write
\begin{equation*}
\begin{split}
&\braket{d\psi_\gamma \otimes d\psi_\gamma}= \sum_{jmn}\braket{d\psi_\gamma |j,m,n}\braket{j,m,n|d\psi_\gamma}\\
&=\sum_{jmn}\int_{SU(2)}dh\,\braket{d\psi_\gamma |j,m,n}\braket{j,m,n|h(A)}\braket{h(A)| d\psi_\gamma}\\
&=\sum_{\substack{jmn\\ j'm'n'}}\int_{SU(2)}dh\,\Bigl(\braket{d\psi_\gamma|j,m,n}\braket{j,m,n|h(A)}\cdot\\
&\qquad\cdot \braket{h(A)|j,m',n'}\braket{j,m',n'|d\psi_\gamma}\Bigr)\\
&=(2j+1)\sum_{\substack{jmn\\ j'm'n'}}\Biggl(\int_{SU(2)}dh\,\overline{D^{(j)}_{mn}(h(A))}D^{(j')}_{m'n'}(h(A))\Biggr)\\
&\qquad d\overline{c}^{j}_{mn}\otimes dc_{mn}^{j}=
\end{split}
\end{equation*}
\begin{equation}\label{f17}
=\sum_{jmn}d\overline{c}^{j}_{mn}\otimes dc_{mn}^{j}\;.
\end{equation}
Similarly, we have
\begin{equation}\label{f18}
\braket{\psi_\gamma|d\psi_\gamma}=\overline{c}^{j}_{mn}\,dc^{j}_{mn}\qquad(\text{with sum over } j,m,n)\;.
\end{equation}

\noindent
Finally, the pull-back to the Hilbert space of the Fubini-Study Hermitian tensor is given by:
\begin{equation}\label{f19}
\begin{split}
\mathcal K_{\mathcal H_\gamma}&=\frac{\braket{d\psi_\gamma\otimes d\psi_\gamma}}{\braket{\psi_\gamma|\psi_\gamma}}-\frac{\braket{d\psi_\gamma|\psi_\gamma}\otimes\braket{\psi_\gamma|d\psi_\gamma}}{\braket{\psi_\gamma|\psi_\gamma}^2}\\
&=\frac{d\overline{c}^{j}_{mn}\otimes dc_{mn}^{j}}{\sum_{mn}|c^{j}_{mn}|^2}-\frac{d\overline{c}^{j}_{mn}\,c^{j}_{mn}\otimes\overline{c}^{j}_{mn}\,dc^{j}_{mn}}{\Bigl(\sum_{mn}|c^{j}_{mn}|^2\Bigr)^2}\;.
\end{split}
\end{equation}
\subsection{Pull-back on orbit submanifolds of quantum states in $\mathcal H_\gamma$}\label{plb}

To pull-back the Hermitian tensor (\ref{f19}) on orbit submanifolds of quantum states we need to understand what are the objects entering the diagram (\ref{gqm89}) in the specific case under examination. 
\\

Let us therefore choose (\ref{f4}) to be our fiducial Wilson line state, i.e.
\begin{equation}\label{f20}
\ket{0}\equiv\ket{\psi_\gamma}=\sum_{jmn}c^{j}_{mn}\ket{j,m,n}\;,
\end{equation}
where we recall from Sec. \ref{GQMa} that $\ket0$ denotes a fiducial fixed state starting from which the orbits under the group action are generated, without any reference to specific properties of the selected state. Since we are considering spin basis states $\ket{j,m,n}$ constructed with the common eigenstates of the operator $J^2$ and one of the $J$'s (say $J_z$), i.e., with a fixed orientation (say the $z$-axis) of the magnetic moments at the endpoints of the link\footnote{We may also consider a more general situation in which we have an additional degree of freedom to take into account a different direction of the magnetic moment. As discussed in \cite{F11}, in this case the basis states are given by $\ket{j,\hat m,\hat n}$, where $\hat m$ simply denotes the new direction ($\sin\theta\cos\varphi, \sin\theta\sin\varphi,\cos\theta$) obtained by rotating the direction $\hat z=(0,0,1)$. This kind of states can be used for istance to account a non-completely precise face matching of polyhedra glued along faces dual to the graph edges which will give some torsion thus providing a generalization of Regge geometries as twisted geometries \cite{F12,F13}.}, the only transformations that we can perform on such states are those generated by the operators $J_1,J_2, J_3$ which have a well-defined action on the basis states. The group $\mathbb G$ acting on $\mathcal H_\gamma$ is thus given by the group $SU(2)$. Therefore, the diagram (\ref{gqm89}) which explains the various levels at which the (co-adjoint) orbit $\mathcal O$ is embedded in the projective Hilbert space $\mathcal R(\mathcal H_\gamma)$ now becomes
\begin{equation}\label{f21}
\xymatrix{
SU(2)\ar[d]_-{U(1)}\ar[r]^-{\phi_0}&S(\mathcal H_\gamma)\ar[d]^-{U(1)}\\
SU(2)\bigl/U(1)\ar[r]^-{\tilde\phi_{0}}\ar[d]_-{\pi_0}&\mathcal R(\mathcal H_\gamma)\\
SU(2)\bigl/\mathbb G_0^{U(1)}\ar[r]^-\cong&\mathcal O\ar[u]_-{i_\mathcal O}
}
\end{equation}
Being $\mathcal H_\gamma$ given by the direct sum of fixed-$j$ Hilbert spaces, i.e.
\begin{equation}\label{wlH}
\mathcal H_\gamma\cong\bigoplus_j\mathcal H_\gamma^{(j)}\cong\bigoplus_j\mathcal V^{(j)}\otimes\mathcal V^{(j)*}\;,
\end{equation}
the SU(2) action on the fiducial state $\ket 0$ is given component-wise. The embedding of the Lie group into $\mathcal{H}_\gamma-\{0\}$ is then realized by means of the action of a spin-$j$-representation on each element of the sum (\ref{wlH}), that is
\begin{equation}\label{f22}
\phi_0\;:\; SU(2)\ni h \longmapsto \ket{h}=U(h)\ket{0}\in\mathcal{H}_\gamma-\{0\}\;,
\end{equation}
where the unitary representation $U:SU(2)\rightarrow Aut(\mathcal H_\gamma)$ is given by
\begin{equation}\label{f23}
U(h)\ket0=\sum_{jmn}c^j_{mn}U^{(j)}(h)\ket{j,m,n}\;,
\end{equation}
and
\begin{equation}
U^{(j)}(t)=e^{iR^{(j)}(X^k)t_k}
\end{equation}
$R^{(j)}(X^k)\equiv J_k$ denoting the set of Hermitian operators which represent the $SU(2)$ generators. Similarly, the corresponding embedding of $\mathbb G\equiv SU(2)$ into the space of rays is given by the co-adjoint action map
\begin{equation}\label{f24}
\tilde\phi_0\;:\;h\longmapsto U^{(j)}(h)\rho_0 U^{(j)\dagger}(h)\;.
\end{equation}
The pull-back of the Hermitian tensor (\ref{f19}) to the co-adjoint orbit starting from a pure fiducial state $\rho_0$ decomposes into a direct sum of the corresponding tensors on each $\mathcal H^{(j)}_\gamma$. We will then focus on a fixed-$j$ block which, according to Eq. (\ref{orbitensor}), will be given by
\begin{equation}\label{f25}
\mathcal K=\mathcal K_{k\ell}\theta^k\otimes\theta^\ell\;,
\end{equation}
with coefficients
\begin{equation}\label{f26}
\mathcal K_{k\ell}=\text{Tr}(\rho_0J_kJ_\ell)-\text{Tr}(\rho_0J_k)\text{Tr}(\rho_0J_\ell)\;.
\end{equation}
Moreover, in the case of a pure state, by using the explicit expression for the fiducial state
\begin{equation}\label{f27}
\rho_0=\frac{\ket 0\bra0}{\braket{0|0}}=\frac{\ket{\psi_{\gamma}}\bra{\psi_\gamma}}{\braket{\psi_\gamma|\psi_\gamma}}\;,
\end{equation}
we find the pulled-back tensor on the corresponding orbits in the Hilbert space:
\begin{equation}\label{f28}
\begin{split}
\mathcal K_{k\ell}&=\frac{\braket{0|J_kJ_\ell|0}}{\braket{0|0}}-\frac{\braket{0|J_k|0}\braket{0|J_\ell|0}}{\braket{0|0}^2}\\
&=\frac{\braket{\psi_\gamma^{(j)}|J_kJ_\ell|\psi_\gamma^{(j)}}}{\braket{\psi_\gamma^{(j)}|\psi_\gamma^{(j)}}}-\frac{\braket{\psi_\gamma^{(j)}|J_k|\psi_\gamma^{(j)}}\braket{\psi_\gamma^{(j)}|J_\ell|\psi_\gamma^{(j)}}}{\braket{\psi_\gamma^{(j)}|\psi_\gamma^{(j)}}^2}\\
&=\braket{J_kJ_\ell}_{\psi_\gamma^{(j)}}-\braket{J_k}_{\psi_\gamma^{(j)}}\braket{J_\ell}_{\psi_\gamma^{(j)}}\;.
\end{split}
\end{equation}
We then see that the Hermitian tensor on the orbits coincides with the covariance matrix of the SU(2) generators. Indeed, starting from the definition of the covariance matrix whose entry in the $k$th row and $\ell$th column is
\begin{equation}\label{f29}
\text{Cov}(J)_{k\ell}=\braket{(J_k-\braket{J_k})(J_\ell-\braket{J_\ell})}\;,
\end{equation}
we have
\begin{align}\label{f30}
&\braket{(J_k-\braket{J_k})(J_\ell-\braket{J_\ell})}=\\ \nonumber
&\qquad=\braket{(J_kJ_\ell-J_k\braket{J_\ell}-\braket{J_k}J_\ell+\braket{J_k}\braket{J_\ell})}\\ \nonumber
&\qquad=\braket{J_kJ_\ell}-\braket{J_k}\braket{J_\ell}-\braket{J_k}\braket{J_\ell}+\braket{J_k}\braket{J_\ell} \\ \nonumber
&\qquad =\braket{J_kJ_\ell}-\braket{J_k}\braket{J_\ell}\;.
\end{align}
The tensor (\ref{f28}) therefore will measure the correlations in the fluctuations of the operators $J$. The non-commutativity of such operators implies that the covariance matrix (\ref{f29}) is not symmetric, but if we remember the decomposition of the Hermitian tensor in its real symmetric and imaginary skewsymmetric part, we find a metric tensor 
\begin{align}\label{f31}
\mathcal K_{(k\ell)}&=\frac{1}{2}\braket{[J_k,J_\ell]_+}_0-\braket{J_k}_0\braket{J_\ell}_0\\ \nonumber
&\equiv\mathbb{R}e\bigl[\braket{(J_k-\braket{J_k})(J_\ell-\braket{J_\ell})}\bigr]\;,
\end{align}
and a symplectic structure 
\begin{equation}\label{f32}
\mathcal K_{[k\ell]}=\mathbb{I}m\biggl(\frac{1}{2}\braket{[J_k,J_\ell]_-}_0\biggr)=\frac{1}{2}\braket{\varepsilon_{k\ell r}J_r}_0\;,
\end{equation}
where we have used the commutation relations $[J_k,J_\ell]_-=i\varepsilon_{k\ell r}J_r$ of the Lie algebra $\mathfrak{su}(2)$.

\subsection{Link as an entangled pair of spherical harmonics}\label{bilink}

The  single link space at fixed $j$ provides the simplest bipartite spin network system, given by the tensor product Hilbert space
\begin{equation}\label{f34}
\mathcal H_\gamma^{(j)}\cong\mathcal V^{(j)}\otimes\mathcal V^{(j)*}\;.
\end{equation}
As we choose an orthonormal basis in the two subspaces, as shown in \eqref{f35}, the single-link state (\ref{f4}) will generally read
\begin{align}\label{f36}
\ket{\psi_\gamma^{(j)}}&=\sum_{mn}c^{(j)}_{mn}\,\ket{j,n}\otimes\ket{j,m}^*\\ \nonumber
&=\sum_{mn}c^{(j)}_{mn}\,\ket{j,n}\otimes\bra{j,m} =\sum_{mn}c^{(j)}_{mn}\,\underset{\ket{j,m,n}}{\underbrace{\ket{j,n}\bra{j,m}}}
\end{align}
namely, as a composite state of two semi-link states analogue to two spherical harmonics $Y^j_m(h)=\braket{h|j,m}$.
Therefore, we can characterize the entanglement of the bipartite system (\ref{f34}) by means of the quantum Fisher tensor description introduced in \ref{GQMb}. Once again, as in \ref{plb}, we restrict our analysis to the case of states with fixed $j$, that is no sum over $j$ in Eq. (\ref{f36}).

According to the diagram (\ref{f21}), we select a fiducial pure state
\begin{equation}\label{f39}
\rho_0\in D^1(\mathcal V^{(j)}\otimes\mathcal V^{(j)*})\cong\mathcal R(\mathcal V^{(j)}\otimes\mathcal V^{(j)*})=\mathcal R(\mathcal H_\gamma^{(j)})\;,
\end{equation}
and then we consider the product representation
\begin{equation}\label{f40}
\phi_0\,:\,\mathbb G\equiv SU(2)\times SU(2)\longrightarrow Aut(\mathcal V^{(j)}\otimes\mathcal V^{(j)*})\;,
\end{equation}
providing the following embedding map
\begin{equation}\label{f41}
\mathbb G\ni g \longmapsto \rho_g=U(g)\rho_0 U^\dagger(g)\in\mathcal R(\mathcal V^{(j)}\otimes\mathcal V^{(j)*})
\end{equation}
with $U(g)=e^{iR(X_k)t}$. Infinitesimal generators $R(X_k)$ are realized as the tensor products between the identity of a subsystem and the spin operators $J_k$ representing the $\mathfrak{su}(2)$ algebra in terms of selfadjoint operators on the Hilbert space $\mathcal V^{(j)}$ (cfr. Eq. (\ref{fisher5})). Thus, according to Sec. \ref{GQMb}, we find that the pull-back of the Hermitian Fisher tensor $\text{Tr}(\rho d\rho\otimes d\rho)$ from $\mathcal R(\mathcal H_{\gamma}^{(j)})=\mathcal R(\mathcal V^{(j)}\otimes\mathcal V^{(j)*})$ to the co-adjoint orbit 
\begin{equation}\label{f42}
\mathcal O_{\rho_0}:=SU(2)\times SU(2)/\mathbb G_{\rho_0}\;,
\end{equation}
where $\mathbb G_{\rho_0}$ is the isotropy group of the fiducial state\footnote{The topology of the orbit will thus depend on the isotropy group of the selected fiducial state. We refer to \cite{F15,F16} for a general discussion.}, decomposes into a symmetric Riemannian and a skewsymmetric (pre-)symplectic component
\begin{equation}\label{f43}
\mathcal K_{k\ell}=\mathcal K_{(k\ell)}+i\mathcal K_{[k\ell]}=\left(\begin{array}{c|c}
A&C\\
\hline
C&B
\end{array}\right)+i\left(
\begin{array}{c|c}
D_A&0\\
\hline
0&D_B
\end{array}
\right)\;,
\end{equation}
with $3\times3$ blocks given by
\begin{equation}\label{f44}
\begin{cases}
A_{ab}=\frac{1}{2}\text{Tr}(\rho_0[J_a,J_b]_+\otimes\mathds 1)-\text{Tr}(\rho_0J_a\otimes\mathds1)\text{Tr}(\rho_0J_b\otimes\mathds1)\\
B_{ab}=\frac{1}{2}\text{Tr}(\rho_0\mathds 1\otimes[J_a,J_b]_+)-\text{Tr}(\rho_0\mathds1\otimes J_a)\text{Tr}(\rho_0\mathds1\otimes J_b)\\
C_{ab}=\text{Tr}(\rho_0J_a\otimes J_b)-\text{Tr}(\rho_0J_a\otimes\mathds1)\text{Tr}(\rho_0\mathds1\otimes J_b)\\
(D_A)_{ab}=\frac{1}{2}\text{Tr}(\rho_0[J_a,J_b]_-\otimes\mathds 1)\\
(D_B)_{ab}=\frac{1}{2}\text{Tr}(\rho_0\mathds 1\otimes[J_a,J_b]_-)
\end{cases}
\end{equation}
Therefore, we see that if $\rho_0$ is maximally entangled, that is the reduced states are maximally mixed
\begin{equation}\label{f46}
\rho_0^{(A)}=\rho_0^{(B)}=\frac{1}{dim\,\mathcal V^{(j)}}\mathds1_{A,B}=\frac{\mathds1_{\mathcal V^{(j)}}}{2j+1}\;,
\end{equation}
then
\begin{equation}\label{f47}
(D_A)_{ab}=\frac{1}{2}\text{Tr}\bigl(\rho_0^{(B)}[J_a,J_b]_-\bigr)\propto\text{Tr}\bigl([J_a,J_b]_-\bigr)=0\;,
\end{equation}
and similarly for $(D_B)_{ab}$. On the other hand, if $\rho_0$ is separable, i.e., $\rho_0=\rho_0^{(A)}\otimes\rho_0^{(B)}$, then
\begin{equation}\label{f48}
\begin{split}
C_{ab}&=\text{Tr}\bigl(\rho_0^{(A)}J_a\otimes\rho_0^{(B)}J_b\bigr)- \\ \nonumber
&\quad - \text{Tr}\bigl(\rho_0^{(A)}J_a\otimes\rho_0^{(B)}\bigr)\text{Tr}\bigl(\rho_0^{(A)}\otimes\rho_0^{(B)}J_b\bigr)\\
&=\text{Tr}\bigl(\rho_0^{(A)}J_a\bigr)\text{Tr}\bigl(\rho_0^{(B)}J_b\bigr)-\\ \nonumber
&\quad - \text{Tr}\bigl(\rho_0^{(A)}J_a\bigr)\underset{1}{\underbrace{\text{Tr}\bigl(\rho_0^{(B)}\bigr)}}\,\underset{1}{\underbrace{\text{Tr}\bigl(\rho_0^{(A)}\bigr)}}\text{Tr}\bigl(\rho_0^{(B)}J_b\bigr)
=0\;.
\end{split}
\end{equation}
Thus, as stated in  Sec. \ref{GQMb},  information about the separability or entanglement of the fiducial state $\rho_0$ is encoded into the different blocks of the pulled-back Hermitian tensor on the orbit of unitarily related states starting from $\rho_0$. Indeed, the vanishing of the symplectic tensor for a maximally entangled state $\rho_0$ corresponds to a vanishing separability while the off-diagonal blocks of the Riemannian tensor are responsible for the entanglement degree of the state $\rho_0$ and allow us to define an associated entanglement monotone $\text{Tr}(C^TC)$ which identifies an entanglement measure geometrically interpreted as a distance between entangled and separable states. As we will discuss later in this work, since we are regarding the link as resulting from the entanglement of semilinks, such entanglement monotone gives us a measure of the existence of the link itself and so of the graph connectivity.

\subsection{Two limiting  cases: maximally entangled and separable states}
In order to visualize the considerations of the previous section, let us focus on the two extreme cases respectively given by a maximally entangled and a separable single link state, and compute explicitly the pull-back of the Hermitian tensor on the orbit having that state as fiducial state. To this aim, we start by considering the Schmidt decomposition \cite{QI6} of the normalized state (\ref{f36}):
\begin{equation}\label{f49}
\ket{\psi_\gamma^{(j)}}=\sum_k\lambda_k\ket{j,k}\otimes\bra{j,k}\;.
\end{equation}
\noindent
In the maximally entangled case all Schmidt coefficients are equal and, according to the normalization condition $\braket{\psi_\gamma^{(j)}|\psi_\gamma^{(j)}}=1$, they are given by:
\begin{equation}\label{f50}
\lambda_k=\frac{1}{\sqrt{2j+1}}\qquad\forall\,k\in[-j,+j]\;,
\end{equation}
\noindent
thus yielding a maximally entangled state
\begin{equation}\label{f51}
\ket{\psi_\gamma^{(j)}}=\frac{1}{\sqrt{2j+1}}\sum_k\ket{j,k}\otimes\bra{j,k}\;,
\end{equation}
\noindent
which is nothing but the gauge-invariant loop state $\ket{\psi_{L}}$. Indeed, such a state corresponds to glue the two endpoints of the link into a bivalent vertex and contract their magnetic moments with an intertwiner provided by the normalized identity in $\mathcal V^{(j)}$, i.e.:

\begin{equation}\label{f52}
\ket{\psi_{L}}=\sum_{k,k'}\frac{\delta_{k,k'}}{\sqrt{2j+1}}\ket{j,k}\otimes\bra{j,k'}\equiv\sum_{k,k'}i_{k,k'}\ket{j,k}\otimes\bra{j,k'}\;.
\end{equation}

\noindent
Therefore, concerning the open single line state regarded as an entangled state of two semilinks, there is a close relationship between maximal entanglement and gauge-invariance. It is actually the gauge-invariance requirement to be responsible for the appearence of entanglement in gluing open spin network states. This is realized by identifying the maximally entangled state (\ref{f51}) with the closed loop state, i.e.

\begin{align}\label{f53}
\mathcal{H}_{max.\,ent.}&\equiv\mathcal{H}_{loop}=\\ \nonumber
&=\text{Inv}_{SU(2)}\bigl[\mathcal V^{(j)}\otimes\mathcal V^{(j)*}\bigr]\subset\mathcal{H}_{\gamma}^{(j)}\;.
\end{align}
However, it should be stressed that the correspondence between gauge invariance and maximal entanglement holds for basis states. Indeed, one can consider gauge invariant superpositions of spin networks, in particular those corresponding to generic cylindrical functions. In this case, the presence of the modes would imply that the states are gauge invariant, but they do not maximize entanglement. Nevertheless, since here we are interested in showing how the GQM machinery introduced in Sec. \ref{GQM} explicitly works, this provides a useful simple example to test the tensorial characterization of entanglement of Sec. \ref{GQMb}. We will then move to the more interesting case of intertwiner entanglement in Sec. \ref{is}.\\

Hence, taking the maximally entangled loop state (\ref{f52}) as our fiducial state, we are interested in the corresponding pulled-back Hermitian tensor on the orbit starting from it. The pure state density matrix $\rho_0\in D^1(\mathcal V^{(j)}\otimes\mathcal V^{(j)*})$ associated with it  is given by
\begin{equation}\label{f54}
\rho_0=\ket{\psi_{L}}\bra{\psi_{L}}=\frac{1}{2j+1}\sum_{k,k'}\bigl(\ket{j,k}\bra{j,k'}\bigr)\otimes\bigl(\ket{j,k'}\bra{j,k}\bigr)\;,
\end{equation}
such that the reduced states are diagonal with eigenvalues exactly given by the square of the Schmidt coefficients, e.g.
\begin{equation}\label{f55}
(\rho_0)_A=\text{Tr}_B(\rho_0)=\frac{1}{2j+1}\sum_k\ket{j,k}\bra{j,k}=\frac{\mathds1_j}{dim\,\mathcal V^{(j)}}\;.
\end{equation}
Hence, by using Eqs. (\ref{f43},\ref{f44}), after lengthy but straightforward calculations, the pull-back of the Hermitian tensor $\mathcal K$ on the orbit $\mathcal O_{\rho_0}$ of Eq. (\ref{f42}) takes  the following form (see the appendix of \cite{tesifabio} for details)
\begin{widetext}
\begin{equation}\label{f56}
\begin{pmatrix}
\frac{1}{3}j(j+1) & 0 & 0 & \frac{1}{3}j(j+1) & 0 & 0 \\
0 & \frac{1}{3}j(j+1) & 0 & 0 & \frac{1}{3}j(j+1) & 0 \\
0 & 0 & \frac{1}{3}j(j+1) & 0 & 0 & \frac{1}{3}j(j+1) \\
\frac{1}{3}j(j+1) & 0 & 0 & \frac{1}{3}j(j+1) & 0 & 0 \\
0 & \frac{1}{3}j(j+1) & 0 & 0 & \frac{1}{3}j(j+1) & 0 \\
0 & 0 & \frac{1}{3}j(j+1) & 0 & 0 & \frac{1}{3}j(j+1) \\
\end{pmatrix}
\end{equation}
\end{widetext}
from which, using the decomposition
$\mathcal K_{k\ell}=\mathcal K_{(k\ell)}+i\mathcal K_{[k\ell]}\;,$
we see that the real symmetric part $\mathcal K_{(k\ell)}$ decomposes in the block-diagonal matrices $A,B$ and the two equal block-off-diagonal matrices $C$, according to
\begin{equation}\label{f58}
\mathcal K_{(k\ell)}=\left(\begin{array}{c|c}
A&C\\
\hline
C&B
\end{array}\right)
\end{equation}
with
\begin{align}\label{f60}
A=B=\begin{pmatrix}
\frac{1}{3}j(j+1) & 0 & 0 \\
0 & \frac{1}{3}j(j+1) & 0 \\
0 & 0 & \frac{1}{3}j(j+1) \\
\end{pmatrix}\;,\\ \nonumber
C=\begin{pmatrix}
\frac{1}{3}j(j+1) & 0 & 0 \\
0 & \frac{1}{3}j(j+1) & 0 \\
0 & 0 & \frac{1}{3}j(j+1) \\
\end{pmatrix}\;,
\end{align}
while the imaginary skewsymmetric part $\mathcal K_{[k\ell]}$\\
\begin{equation}\label{61}
\mathcal K_{[k\ell]}=\left(\begin{array}{c|c}
D_A & 0 \\
\hline
0 & D_B
\end{array}\right)\;\;\text{with}\;\;D_A=D_B=\begin{pmatrix}
0 & 0 & 0 \\
0 & 0 & 0 \\
0 & 0 & 0 \\
\end{pmatrix}\;,
\end{equation}
gives a vanishing symplectic structure, as expected for the maximally entangled case. Moreover, by using the off-diagonal blocks (\ref{f60}) of the Riemannian symmetric part, we have
\begin{equation}\label{f63}
\text{Tr}(C^TC)=\sum_{a,b=1}^3\,C_{ab}^2=\frac{1}{3}[j(j+1)]^2\;.
\end{equation}
The associated entanglement monotone (\ref{fisher21}) is given by
\begin{equation}\label{2f63}
\begin{split}
\mathcal E&=\frac{(2j+1)^2}{4[(2j+1)^2-1]}\,\frac{1}{3}[j(j+1)]^2\\
&=\frac{1}{48}j(j+1)[4j(j+1)+1]\;,
\end{split}
\end{equation}
while the geometric distance $\Sigma$ defined in (\ref{geomdist}) reads as
\begin{equation}
\Sigma=\frac{1}{3(2j+1)^4}[j(j+1)]^2\;.
\end{equation}
Let us notice that the entanglement measure $\mathcal E$ depends only on the area eigenvalue $j(j+1)$. Moreover, for large $j$, Eqs. (\ref{f63}) and (\ref{2f63}) coincide up to a numerical factor, while in the same limit the geometric distance $\Sigma$ is independent of $j$ as it should.

On the other extreme, if we consider a separable fiducial state, the two spin states do not talk with each other and may have in general different spins, i.e.:
\begin{equation}\label{f64}
\ket{0}=\ket{j_1,k_1}\otimes\bra{j_2,k_2}\;.
\end{equation}
The corresponding pure state density matrix is given by
\begin{equation}\label{f65}
\rho_0=\rho_0^{(A)}\otimes\rho_0^{(B)}=\bigl(\ket{j_1,k_1}\bra{j_1,k_1}\bigr)\otimes\bigl(\ket{j_2,k_2}\bra{j_2,k_2}\bigr)\;.
\end{equation}
Hence, the pull-back of the Hermitian tensor $\mathcal K$ on the orbit $\mathcal O_{\rho_0}$ will take the following form \cite{tesifabio}
\begin{widetext}
\begin{equation}\label{f66}
\begin{varpmatrix}[\scriptsize]
\frac{1}{2}[j_1(j_1+1)-k_1^2] & \frac{i}{2}k_1 & 0 & 0 & 0 & 0 \\
-\frac{i}{2}k_1 & \frac{1}{2}[j_1(j_1+1)-k_1^2] & 0 & 0 & 0 & 0 \\
0 & 0 & k_1(k_1-k_2) & 0 & 0 & 0 \\
0 & 0 & 0 & \frac{1}{2}[j_2(j_2+1)-k_2^2] & \frac{i}{2}k_2 & 0 \\
0 & 0 & 0 & -\frac{i}{2}k_2 & \frac{1}{2}[j_2(j_2+1)-k_2^2] & 0 \\
0 & 0 & 0 & 0 & 0 & k_2(k_2-k_1) \\
\end{varpmatrix}
\end{equation}
\end{widetext}
from which we see that, as expected for the separable case, we have vanishing off-diagonal block matrices $C$ and a direct sum
\begin{widetext}
\begin{equation}\label{f67}
\underset{\mathcal K_A}{\underbrace{\begin{varpmatrix}[\scriptsize]
\frac{1}{2}[j_1(j_1+1)-k_1^2] & \frac{i}{2}k_1 & 0 \\
-\frac{i}{2}k_1 & \frac{1}{2}[j_1(j_1+1)-k_1^2] & 0 \\
0 & 0 & k_1(k_1-k_2) \\
\end{varpmatrix}}}\oplus\underset{\mathcal K_B}{\underbrace{\begin{varpmatrix}[\scriptsize]
\frac{1}{2}[j_2(j_2+1)-k_2^2] & \frac{i}{2}k_2 & 0 \\
-\frac{i}{2}k_2 & \frac{1}{2}[j_2(j_2+1)-k_2^2] & 0 \\
0 & 0 & k_2(k_2-k_1) \\
\end{varpmatrix}}}
\end{equation}
\end{widetext}
of two decoupled Hermitian tensors $\mathcal K_A$ and $\mathcal K_B$ one for each subsystem. Moreover, a further decomposition of the Hermitian tensor (\ref{f66}) as
$\, \mathcal K_{k\ell}=\mathcal K_{(k\ell)}+i\mathcal K_{[k\ell]}\;, $
gives a symmetric real part
\begin{equation}\label{f69}
\mathcal K_{(k\ell)}=\left(\begin{array}{c|c}
A&C\\
\hline
C&B
\end{array}\right)
\end{equation}
with
\begin{equation}\label{f70}
A=\begin{pmatrix}
\frac{1}{2}[j_1(j_1+1)-k_1^2] & 0 & 0 \\
0 & \frac{1}{2}[j_1(j_1+1)-k_1^2] & 0 \\
0 & 0 & k_1(k_1-k_2) \\
\end{pmatrix}\;,
\end{equation}
\begin{align}\label{f70b}
&B=\begin{pmatrix}
\frac{1}{2}[j_2(j_2+1)-k_2^2] & 0 & 0 \\
0 & \frac{1}{2}[j_2(j_2+1)-k_2^2] & 0 \\
0 & 0 & k_2(k_2-k_1) \\
\end{pmatrix}\;, \\ \nonumber
&C=\begin{pmatrix}
0 & 0 & 0 \\
0 & 0 & 0 \\
0 & 0 & 0 \\
\end{pmatrix}\;,
\end{align}
and an imaginary skewsymmetric part
\begin{equation}\label{f72}
\mathcal K_{[k\ell]}=\left(\begin{array}{c|c}
D_A & 0 \\
\hline
0 & D_B
\end{array}\right)
\end{equation}
with
\begin{equation}\label{f73}
D_A=\begin{pmatrix}
0 & \frac{1}{2}k_1 & 0 \\
-\frac{1}{2}k_1 & 0 & 0 \\
0 & 0 & 0 \\
\end{pmatrix}\,\,,\qquad D_B=\begin{pmatrix}
0 & \frac{1}{2}k_2 & 0 \\
-\frac{1}{2}k_2 & 0 & 0 \\
0 & 0 & 0 \\
\end{pmatrix}\;.
\end{equation}
Finally, we have
\begin{equation}\label{f74}
\text{Tr}(C^TC)=0\;,
\end{equation}
i.e., coherently with its interpretation as a distance from the separable state, the entanglement measure associated with the block-off-diagonal matrices $C$ is zero in the unentangled case.

\section{Gluing links by entanglement}\label{2wl}
Let us proceed a little step further with respect to what shown in the previous section, and consider now the description of the entanglement resulting from the gluing of two lines into one. The bipartite Hilbert space is given by two copies of a single link Hilbert space with fixed but different spin labels, i.e.
\begin{equation}\label{f75}
\mathcal H=\mathcal H_{\gamma_1}^{(j_1)}\otimes\mathcal H_{\gamma_2}^{(j_2)}\;,
\end{equation}
and the fiducial state is chosen to be a single line state coming from the gluing of two other links, thus  admitting the following expression
\begin{equation}\label{f76}
\ket{0}\equiv\ket{\psi_\gamma}=\frac{1}{\sqrt{2j+1}}\sum_{m,n,k,\ell}c_{mn}\ket{j,m,k}\otimes\ket{j,\ell,n}\delta_{k,\ell}\;.
\end{equation}
The local $SU(2)$ gauge-invariance requirement at the gluing point $v\equiv\gamma_1(1)=\gamma_2(0)$, implemented by the bivalent intertwiner $\delta_{k,\ell}/\sqrt{2j+1}$ contracting the magnetic numbers of the glued endpoints, forces the two spins to be equal, i.e. $j_1=j_2=j$.% (see Fig. \ref{gluedlink}).
%\begin{figure}[t!]
%\centering
%\includegraphics[scale=0.30]{fig3.jpg}
%\caption{Gluing links by entanglement}
%\label{gluedlink}
%\end{figure}
In other words, $\ket{0}$ is a locally $SU(2)$-invariant state in $\mathcal H$, that is
\begin{equation}\label{f77}
\ket{0}\;\in\;\mathcal H_\gamma^{(j)}\subset\mathcal H\,,\qquad\gamma=\gamma_1\circ\gamma_2\;.
\end{equation}
However, in order to compute an entanglement measure  which can be interpreted as the distance of our fiducial state from the separable one, we need to consider the action $\phi$ of a Lie group $\mathbb G$ on $\mathcal H$ and not only on the gauge-reduced level $\tilde\phi:\mathbb G/SU(2)\rightarrow\mathcal H_\gamma^{(j)}$. Therefore, the underlying scheme of the construction of the pulled-back Hermitian tensor on the orbit of states with fixed amount of entanglement will be given by the following diagram
\begin{equation}\label{f78}
\xymatrix{
\mathbb G\ar[dd]_-{SU(2)}\ar[r]^-{\phi_0} & \mathcal H_{\gamma_1}^{(j_1)}\otimes\mathcal H_{\gamma_2}^{(j_2)}\ar[dd]^-{SU(2)}\\
\ar@{}[r]^-{``\text{gluing}"} & \\
\mathbb G\bigl/SU(2)\ar[r]^-{\tilde\phi_{0}}\ar[d]_-{\pi_0} & \mathcal H_{\gamma_1\circ\gamma_2}^{(j)}\\
\mathbb G\bigl/\mathbb G_0\ar[r]^-\cong & \mathcal O\ar[u]_-{i_\mathcal O}
}
\end{equation}
We recall that the group $\mathbb G$ is a group of local unitary transformations which as such do not modify the degree of entanglement along the orbit starting at the selected fiducial state. In the specific case under consideration, the group $\mathbb G$ is $SU(2)$ and its action on the bipartite Hilbert space (\ref{f75}) is realized through a product representation
\begin{equation}\label{f79}
U(\mathcal H)=U(\mathcal H_{\gamma_1})\otimes U(\mathcal H_{\gamma_2})\;,
\end{equation}
whose infinitesimal generators are given by the SU(2)-generators tensored by the identity of one of the subsystems. Indeed, each subsystem Hilbert space reads as
\begin{equation}\label{f80}
\mathcal H_{\gamma_i}^{(j)}\cong\mathcal V^{(j_i)}\otimes\mathcal V^{(j_i)*}\qquad\quad (i=1,2)\;,
\end{equation}
and so the bipartite Hilbert space (\ref{f75}) can be regarded as
\begin{equation}\label{f81}
\mathcal H\cong(\mathcal V^{(j_1)}\otimes\mathcal V^{(j_1)*})\otimes(\mathcal V^{(j_2)}\otimes\mathcal V^{(j_2)*})\;.
\end{equation}
The gluing operation $\gamma=\gamma_1\circ\gamma_2$ corresponds to select the subspace
\begin{equation}\label{f82}
\mathcal V^{(j_1)}\otimes\text{Inv}_{SU(2)}\bigl[\mathcal V^{(j_1)}\otimes\mathcal V^{(j_2)*}\bigr]\otimes\mathcal V^{(j_2)}\;\subset\;\mathcal H\,,
\end{equation}
which reduces to
\begin{equation}\label{f83}
\mathcal V^{(j)}\otimes\mathcal V^{(j)*}\cong\mathcal H_\gamma^{(j)}\,,\qquad j=j_1=j_2
\end{equation}
since, according to the \textit{Schur's lemma} \cite{LQG31}, when we have only two spin representations the invariant bivalent intertwining operator $\mathcal V^{(j_1)}\rightarrow\mathcal V^{(j_2)}$ is either proportional to the identity if $j_1=j_2$ or zero if $j_1\neq j_2$, i.e., the invariant subspace is trivial.\\We are thus brought back to the situation of the previous section. The pulled-back Hermitian tensor $\mathcal K$ is again given by  the pull-back of (\ref{f44}) with a fiducial state now given by (\ref{f76}) and the spin operators $J$ act non-trivially only at the free endpoints of the resulting new link. Hence, there is no need to repeat our calculations and we only notice that, coherently with the general considerations of Section \ref{GQM}, we have:
\begin{itemize}
\item For  the skewsymmetric part:
\begin{align}\label{f84}
%\begin{split}
&(D_A)_{ab}=\braket{0|[J_a,J_b]_-\otimes\mathds1|0}\\ \nonumber
&=\frac{1}{2j+1}\sum_{\substack{mn\ell \\ m'n'\ell'}}\overline{c_{m'n'}}c_{mn}\braket{j,m',\ell'|[J_a,J_b]_-|j,m,\ell}\delta_{\ell',\ell}\delta_{n',n}\\ \nonumber
& =\frac{1}{2j+1}\sum_{mn\ell m'}\overline{c_{m'n}}c_{mn}\braket{j,m',\ell|[J_a,J_b]_-|j,m,\ell}\\ \nonumber
&=\sum_{mnm'}\overline{c_{m'n}}c_{mn}\braket{j,m'|[J_a,J_b]_-|j,m}
%\end{split}
\end{align}
being  $\braket{j,\ell|j,\ell}=1$. We then see that when the fiducial state is maximally entangled, i.e., all the Schmidt coefficients are equal, we end up with the trace of the commutator which is zero.
\item By similar arguments, when $\ket0$ is separable, we see that the block-off-diagonal matrices of the symmetric part vanish:
\begin{align}\label{f85}
C_{ab}&=\braket{0|J_a\otimes J_b|0}-\braket{0|J_a\otimes\mathds1|0}\braket{0|\mathds1\otimes J_b|0}\\ \nonumber
&=\braket{J_a}\braket{J_b}-\braket{J_a}\braket{J_b}=0\;.
\end{align}
\end{itemize}
Let us stress again that here for simplicity we content our analysis to fixed spin $j$ labelling the $SU(2)$ irreducible representations associated to the links. Nevertheless, we could also consider the action of a product unitary representation on the full Hilbert spaces $\mathcal H_{\gamma_1}\otimes\mathcal H_{\gamma_2}$ where, as noticed in section \ref{plb}, the action on each subsystem will be realized in terms of the action on each element of the direct sum $\mathcal H_{\gamma_i}\cong\bigoplus_{j_i}\mathcal H_{\gamma_i}^{(j_i)}, i=1,2$. Obviously, the resulting expression of the Fisher tensor will be more complicated in this case but it will acquire the following structure which takes into account different kinds of correlations between the subsystems. The diagonal blocks will refer to different subspaces in the direct sum characterized by a different value of the spin labels. Within each block, the off-diagonal terms will therefore encode correlations between fixed-spin subspaces. On the other hand, the off-diagonal blocks of the full tensor will encode possible correlations between different spin configurations. This would allow to extend the study of correlations also to the gluing of full Wilson line states and, more generally, of open spin network states. We will discuss such cases elsewhere.

\section{unfolding intertwiner into a pair of entangled qdits}\label{is}
We shall now extend our tensorial approach to a class of  quantum spin network states with a more involved nonlocal correlation structure.  

In particular we look at the most concise description for a bounded region of quantum space in the framework of LQG spin networks, given by polyhedra dual to a superpositions of $n$-valent intertwiner states (see Appendix \ref{spin} for a detailed definition of intertwiner), with support on a single vertex graphs with $n$ links.
In absence of non-trivial internal curvature degrees of freedom \cite{lt1}, one can dually think of such region as a portion of 3d space flatly embedded in 4d. The irreducible representations  carried by the open edges are dual to boundary patches comprising the quantum surface of the convex flat polyhedron dual to the intertwiner. 

In particular, we can consider an ideal partition of the surface boundary, by dividing $n$ into two sets $n_A=(\{j_e\}|e \in \partial A)$ and $n_B=(\{j_e\}|e \in \partial B)$
and think of the intertwiner state as a \emph{bipartite} quantum mechanical system constrained by an overall $SU(2)$ gauge invariance. 

The full intertwiner Hilbert space, in the notation given \eqref{qg77}, reads
\begin{eqnarray}
\mathcal H_V&=&\bigoplus_{\{j_e\}}\bigl(\text{Inv}_{SU(2)}\bigl[\bigotimes_{e\in v}\mathcal V^{j_e}\bigr]\bigr)\\ \nonumber
&\cong& L^2\left(SU(2)^L/SU(2)\right)
\end{eqnarray}
This space has infinite dimension, due to the direct sum structure over the spin-$J$ representations.  Therefore, as for the bipartite link in \ref{bilink}, we choose to further restrict our analysis to the finite dimensional case of the single intertwiner space at fixed $\{j_e\}$,
\begin{eqnarray}\label{into}
\mathcal{H}_v^{\{j_e\}} =\text{Inv}_{SU(2)}\bigl[\bigotimes_{ e \in v}\mathcal V^{j_e}\bigr]
\end{eqnarray}
This is still considerably more involved than a single link. However, we must point out that, from the point of view of both LQG and GFT, it remains a drastic truncation of the set of (kinematical) degrees of freedom, and of the possibly relevant states, particularly from the point of view of a reconstruction of an approximate continuum spacetime and geometry. Indeed, it implies a truncation to a finite set of degrees of freedom loosing the functional aspects of the theory.\footnote{It should also be noted, though, that most work in the LQG context is limited to such fixed-graph computations, due to the complexity of the more general case. In particular, the following analysis is framed in a very similar way as the one in \cite{F3}.}
\subsection{The bipartite system}

%
%We restrict our analysis to the case of a region of quantum space without non-trivial internal curvature degrees of freedom \cite{lt1}. The absence of loops carrying curvature excitations at the reduced, gauge-fixed vertex allows us to dually think of such region as a portion of 3d space flatly embedded in 4d.
%In this case, the flower graph reduces to a $n$-valent intertwiner (see Fig. \ref{flower1}). The irreducible representations  carried by the open edges are dual to boundary patches comprising the quantum surface of the convex flat polyhedron dual to the intertwiner. In particular, we want to divide $n$ in the two sets $n_A=(\{j_e\}|e \in \partial A)$ and $n_B=(\{j_e\}|e \in \partial B)$.

We consider then a generic single intertwiner space as a \emph{bipartite} quantum mechanical system constrained by an overall $SU(2)$ gauge invariance. 
To the given  separation of the boundary degrees of freedom, we can associate two boundary spaces, respectively defined by the tensor product of the $SU(2)$ irreps labelling the edges
\begin{align}\label{tenso}
\mathcal{H}_A\equiv \bigotimes_{e \in \partial A}\mathcal{V}^{j_e} \qquad\text{and}\quad \mathcal{H}_B\equiv \bigotimes_{e \in \partial B} \mathcal{V}^{j_e}. 
\end{align}
In the intertwiner, the two boundary sets of degrees of freedom are constrained by the gauge invariance at the vertex. Indeed, we can rewrite \eqref{into} as
\begin{align} \label{unfolded1}
\mathcal{H}_v^{\{j_e\}}= \text{Inv}_{SU(2)}\,[ \mathcal{H}_A \otimes  \mathcal{H}_B]
\end{align}
Such a constraint is at the root of the non-local quantum correlations among the edges.

%Now, for what concerns our analysis, we further simplify the problem by considering the case of a single intertwiner graph space, namely we work at fixed $\{j_e\}$, avoiding the full sum $\bigoplus_{\{j_e\}}$ in the above \eqref{unfolded1}. This is another strong simplification, since it implies a further reduction of the degrees of freedom being considered, which were already truncated to a  finite number by fixing a graph $\Gamma$.

Let us then proceed by taking the decomposition of the two tensor product spaces in \eqref{tenso} in direct sums of irreducible representations: we write each subspace $\mathcal{H}_{A,B}$ as $\bigoplus_k \mathcal{V}^k \otimes \mathcal{D}_k $, where $\mathcal{V}^k$ is a $(2k + 1)$-dimensional re-coupled spin-$k$ irrep of $SU(2)$, coming with its degeneracy space $\mathcal{D}_k$.
In these terms, we write
\begin{align} \nonumber \label{unfolded2} 
&\mathcal{H}_{v}^{\{j_e\}}= \text{Inv}_{SU(2)}\,[ \mathcal{H}_A \otimes  \mathcal{H}_B]\\ \nonumber
&=\text{Inv}_{SU(2)}\left[ \bigoplus_{k,y} (\mathcal{V}_{A}^k \otimes \,^{\{j_{e \in \partial A}\}}\mathcal{D}_k)\otimes(\mathcal{V}_{B}^y \otimes \,^{\{j_{e \in \partial B}\}}\mathcal{D}_y) \right]\\ \nonumber
%\end{align}
%\begin{align} \nonumber 
&=  \bigoplus_{k,y} \delta_{k,y}\left[  (\mathcal{V}^k_A \otimes \mathcal{V}^y_B)\otimes( \,^{\{j_{e \in\partial A}\}} \mathcal{D}_k\otimes \,^{\{j_{e \in\partial B}\}}\mathcal{D}_y) \right]\\ \nonumber
& =\bigoplus_{k} (\mathcal{V}^k_A \otimes  \,^{\{j_{e \in\partial A}\}} \mathcal{D}_k )\otimes(\mathcal{V}^k_B\otimes \,^{\{j_{e \in\partial B}\}}\mathcal{D}_k)\\ 
&= \bigoplus_{k} \,^{\{j_{e \in\partial A}\}}\mathcal{H}_A^{(k)} \otimes \,^{\{j_{e \in\partial B}\}}\mathcal{H}_B^{(k)},
\end{align}
where, by gauge invariance, the two re-coupled spin irreps appearing in the second line are tensored to form a trivial representation for each $k$.
Consistently, the dimension of the single intertwiner space reduces to
\begin{align}\nonumber \label{dimension of R}
N_v &\equiv  \text{dim}( \mathcal{H}_{v}^{\{j_e\}})
=\text{dim} \left[\bigoplus_{k} (\mathcal{V}^k_A \otimes \mathcal{V}^k_B )\otimes( \mathcal{D}_k^A \otimes \mathcal{D}_k^B) \right] \\
&=\sum_k \,\text{dim} \, \mathcal{D}_k^A \cdot \text{dim} \,\mathcal{D}_k^B
= \sum_k N^{k}_{A} \,N^{k}_{B}.
\end{align}
where we define $N^{k}_{A,B}\equiv d_k^{(n_{A,B})}=\text{dim} \,\mathcal{D}_k^{(A,B)} $, the dimensions of the degeneracy spaces. 

Starting from the decomposition in \eqref{unfolded2}, a convenient basis in the two subsystems $A$ and $B$ is labeled by three numbers, respectively $|k,m,\alpha_k\rangle$ and $|k,m,\beta_k\rangle$, with $\alpha_k, \beta_k$ giving the number of the different irreducible representations $\mathcal{V}^k_{A,B}$ for given $k$ \cite{etbe}. A basis for the single intertwiner space is then written as
\begin{align} \label{basis}
| k, \alpha_k , \beta_k \rangle = \sum_{m=-k}^k \frac{(-1)^{k-m}}{\sqrt{2k+1}}  |k, -m, \alpha_k \rangle_A\, \otimes  | k, m, \beta_k \rangle_B
\end{align}
Given the peculiar tensor structure of the unfolded intertwiner space, each basis state can be represented as a tensor product state on three subspaces \cite{etbe},
\begin{align} \label{biortho}
|k, \alpha_k , \beta_k \rangle \equiv  |{k} \rangle_{\mathcal{V}_A^k\otimes \mathcal{V}_B^k} \otimes |{\alpha_k}\rangle_{\mathcal{D}_k^A} \otimes |{\beta_k}\rangle_{\mathcal{D}_k^B}
\end{align}
where the generic $|\zeta_k\rangle$ labels a basis vector of $\mathcal{D}_k$, with $\zeta_k$ running from 1 to $N_k=\text{dim} \,\mathcal{D}_k$.
Therefore, a generic state vector in $\mathcal{H}_{v}^{\{j_e\}}$ is given by a superposition of product basis states of the three subspaces, 
\begin{align} \label{pure}
|\psi_{v}\rangle&= \sum_{\substack{k,\\ \alpha_k, \beta_k}} c_{k,\alpha_k, \beta_k}^{(\{j_e\})}\, |k \rangle_{\mathcal{V}_A^k\otimes\mathcal{V}_B^k} \otimes |{\alpha_k}\rangle_{\mathcal{D}_k^A} \otimes |{\beta_k}\rangle_{\mathcal{D}_k^B}\;.
\end{align}
In particular, we focus our analysis on a specific class of states, generically written as
\begin{align} \label{pure2}
|\psi_v\rangle_k&= \sum_{\alpha_k, \beta_k} c_{\alpha_k, \beta_k}^{(k,\{j_e\})}\, |k \rangle_{\mathcal{V}_A^k\otimes\mathcal{V}_B^k} \otimes |{\alpha_k}\rangle_{\mathcal{D}_k^A} \otimes |{\beta_k}\rangle_{\mathcal{D}_k^B},
\end{align}
with no sum over $k$. This means that we discard any quantum correlation among the tensored irreps space and the degeneracy spaces and we look at the entanglement induced by correlations among the degeneracy spaces only. This partially reduces the complexity of the problem. 

By fixing the virtual link spin $k$, the unfolded intertwiner space (\ref{unfolded2}) admits the following tensor product structure
\begin{equation}\label{Hfixedk}
\begin{split}
\mathcal H_v^{k,\{j_e\}}&=\,^{\{j_{e \in\partial A}\}}\mathcal{H}_A^{(k)} \otimes \,^{\{j_{e \in\partial B}\}}\mathcal{H}_B^{(k)}\\ \nonumber
&=(\mathcal V_A^k\otimes\,^{\{j_{e \in\partial A}\}}\mathcal{D}_k)\otimes(\mathcal V_B^k\otimes\,^{\{j_{e \in\partial B}\}}\mathcal{D}_k)\\
&=(\mathcal V_A^k\otimes\mathcal V_B^k)\otimes\mathcal D_k^A\otimes\mathcal D_k^B\;.
\end{split}
\end{equation}
A generic pure state density matrix $\rho_v=\ket{\psi_v}\bra{\psi_v}\in D^1(\mathcal H_v^{k,\{j_e\}})$ will then have the following simplified form
\begin{equation}\label{fixedkstate}
\rho_v\equiv\ket{\psi_v}\bra{\psi_v}=\rho^{(k)}\otimes\rho_{AB}^{(k)}\;,
\end{equation}
with
\begin{equation}\label{kstate}
\rho^{(k)}=\ket{k}\bra{k}\;\in\;D^1(\mathcal V_A^k\otimes\mathcal V_B^k)\;,
\end{equation}
and
\begin{align}\label{Dstate}
\rho_{AB}^{(k)}=\sum_{\substack{\alpha_k, \beta_k\\\alpha'_k,\beta'_k}} c_{\alpha_k, \beta_k}^{(k,\{j_e\})}\,\overline{c}_{\alpha'_k, \beta'_k}^{(k,\{j_e\})}\,|{\alpha_k}\rangle\langle{\alpha'_k}|\otimes |{\beta_k}\rangle\langle{\beta'_k}|\;,
\end{align}
in $D^1(\mathcal D_k^A\otimes\mathcal D_k^B)$. With respect to the total space $\mathcal H_v^{k,\{j_e\}}$ at fixed $k$, the pure state $\rho_v$ is separable and can be factorized into a tensor product between a pure state $\rho^{(k)}$ involving only the spin $k$ irreps and a pure state $\rho_{AB}^{(k)}$ possibly entangling the degeneracy spaces. Since there are no correlations among the tensored irreps space and the degeneracy spaces, we can trace out $\rho^{(k)}$ and focus our attention only on the state $\rho_{AB}^{(k)}$ over the degeneracy spaces.

In these terms, for each $k$, we can effectively treat the intertwiner state as an entangled pure state on the bipartite degeneracy space $\mathcal{D}_k^A \otimes \mathcal{D}_k^B$. In this sense, we can describe the degeneracy structure of the unfolded intertwiner state as a couple of entangled quantum $N$-level systems with number of levels provided by the degeneracy factors $N_X^k=\text{dim\,}\mathcal D_k^X\equiv d_k^{(n_X)}$, $X=A,B$, respectively.\\
%
%Again we are interested in studying the quantum correlations between the two non-adjacent spin network regions $A$ and $B$, which we may think to be induced by the spin network state outside $A$ and $B$, i.e., by the outside space geometry defined by the complementary region $R$ to $A\cup B$ within $\Gamma$. We expect such correlations to be related to a notion of distance between $A$ and $B$, that transcends the direct connectivity information encoded in the links connecting the two regions.
%
Along the lines of the derivation given in Section \ref{GQM}, we now proceed in investigating the correlation structure of the graph by focusing on the tensorial structures intrinsically defined on the degeneracy spaces of the unfolded intertwiner state.

\subsection{Quantum Fisher Tensor on $D^1(\mathcal D_k^A\otimes\mathcal D_k^B)$}
Our goal consists now in computing the full quantum Fisher tensor on the orbit submanifolds identified by the fiducial bipartite state (\ref{Dstate}).

In order to calculate the entanglement monotone in (\ref{fisher21}) explicitly, we need to put the block-coefficient matrices (\ref{fisher19}), (\ref{fisher199}) of the Hermitian tensor $\mathcal K$ in a more manageable form. To this aim, we use the \textit{stardard} (or \textit{natural}) basis over complex numbers for the $\mathfrak u(N)$ Lie algebras associated with the two subsystems, that is 
\begin{equation}
\sigma_a^{(A)}\longmapsto\tau_{aa'}^{(A)}\equiv\ket{a_k}\bra{a_k'},\quad\sigma_b^{(B)}\longmapsto\tau_{bb'}^{(B)}\equiv\ket{b_k}\bra{b_k'}
\end{equation}
with $a_k,a_k'=1,\dots,N^k_A\equiv d^{(n_A)}_k$ and $b_k,b_k'=1,\dots,N^k_B\equiv d^{(n_B)}_k$ such that
\begin{equation}\label{fisher23}
\left(\tau_{aa'}^{(A)}\right)_{cc'}=\delta_{ac}\delta_{a'c'}\;,
\end{equation}
and similarly for the $\tau^{(B)}$'s. Such a change of basis essentially amounts to replace the $\sigma$'s with the $\tau$'s in the expressions (\ref{fisher19}), (\ref{fisher199}). Indeed, the Lie algebra-valued left-invariant 1-form $U^{-1}dU$ can still be decomposed as in Eq. (\ref{fisher7}), say
\begin{align}\label{fisher24}
U^{-1}dU&=U_A^{-1}d_AU_A\otimes\mathds1_B+\mathds1_A\otimes U_B^{-1}d_BU_B\\ \nonumber
&=i\tau_{aa'}^{(A)}\theta^{aa'}_A\otimes\mathds1_B+\mathds1_A\otimes i\tau_{bb'}^{(B)}\theta^{bb'}_B\;,
\end{align}
where, for each subsystem $X=A,B$, the $\tau^{(X)}$ can be expressed as linear combinations of the infinitesimal generators $\sigma^{(X)}$ with complex coefficients and, correspondingly, also the new bases of left-invariant 1-forms (i.e., $\{\theta^{aa'}_A\}$ and $\{\theta^{bb'}_B\}$) are now given in terms of $\mathbb C-$linear combinations of the previous ones\footnote{For instance, in the $U(2)$ case we have:
$$
\sigma_0=\begin{pmatrix}1&0\\0&1\end{pmatrix},\quad\sigma_1=\begin{pmatrix}0&1\\1&0\end{pmatrix},\quad\sigma_2=\begin{pmatrix}0&-i\\i&0\end{pmatrix},\quad\sigma_3=\begin{pmatrix}1&0\\0&-1\end{pmatrix}
$$
$$
\tau_{11}=\begin{pmatrix}1&0\\0&0\end{pmatrix},\quad\tau_{12}=\begin{pmatrix}0&1\\0&0\end{pmatrix},\quad\tau_{21}=\begin{pmatrix}0&0\\1&0\end{pmatrix},\quad\tau_{22}=\begin{pmatrix}0&0\\0&1\end{pmatrix}\;.
$$
Therefore
$$
\sigma_0=\tau_{11}+\tau_{22},\quad\sigma_1=\tau_{12}+\tau_{21},\quad\sigma_2=i(\tau_{21}-\tau_{12}),\quad\sigma_3=\tau_{11}-\tau_{22}
$$
from which, by imposing that $U^{-1}dU=i\sigma_k\theta^k=i\tau_{kk'}\theta^{kk'}$, it is easy to see that
$$
\theta^{11}=\theta^0+\theta^3,\quad\theta^{12}=\theta^1-i\theta^2,\quad\theta^{21}=\theta^1+i\theta^2,\quad\theta^{22}=\theta^0-\theta^3\;.
$$
}.
With this choice of basis, the block-coefficient matrices (\ref{fisher19}), (\ref{fisher199}) then become:
\begin{widetext}
\begin{equation}\label{fisher25}
\begin{cases}
\mathcal K_{\substack{(aa')\\(bb')}}^{(A)}=\frac{1}{2}\text{Tr}\left(\rho_0\bigl[\tau_{aa'}^{(A)},\tau_{bb'}^{(A)}\bigr]_+\otimes\mathds1_B\right)-\text{Tr}\left(\rho_0\tau_{aa'}^{(A)}\otimes\mathds1_B\right)\text{Tr}\left(\rho_0\tau_{bb'}^{(A)}\otimes\mathds1_B\right)
\\
\mathcal K_{\substack{(aa')\\(bb')}}^{(B)}=\frac{1}{2}\text{Tr}\left(\rho_0\mathds1_A\otimes\bigl[\tau_{aa'}^{(B)},\tau_{bb'}^{(B)}\bigr]_+\right)-\text{Tr}\left(\rho_0\mathds1_A\otimes\tau_{aa'}^{(B)}\right)\text{Tr}\left(\rho_0\mathds1_A\otimes\tau_{bb'}^{(B)}\right)
\\
\mathcal K_{\substack{(aa')\\(bb')}}^{(AB)}=\text{Tr}\left(\rho_0\tau_{aa'}^{(A)}\otimes\tau_{bb'}^{(B)}\right)-\text{Tr}\left(\rho_0\tau_{aa'}^{(A)}\otimes\mathds1_B\right)\text{Tr}\left(\rho_0\mathds1_A\otimes\tau_{bb'}^{(B)}\right)
\end{cases}
\end{equation}

\begin{equation}\label{fisher25b}
\begin{cases}\mathcal K_{\substack{(aa')\\(bb')}}^{(BA)}=\text{Tr}\left(\rho_0\tau_{aa'}^{(B)}\otimes\tau_{bb'}^{(A)}\right)-\text{Tr}\left(\rho_0\tau_{aa'}^{(B)}\otimes\mathds1_A\right)\text{Tr}\left(\rho_0\mathds1_B\otimes\tau_{bb'}^{(A)}\right)
\\
\mathcal K_{\substack{[aa']\\ [bb']}}^{(A)}=\frac{1}{2}\text{Tr}\left(\rho_0\bigl[\tau_{aa'}^{(A)},\tau_{bb'}^{(A)}\bigr]_-\otimes\mathds1_B\right)
\\
\mathcal K_{\substack{[aa']\\ [bb']}}^{(B)}=\frac{1}{2}\text{Tr}\left(\rho_0\mathds1_A\otimes\bigl[\tau_{aa'}^{(B)},\tau_{bb'}^{(B)}\bigr]_-\right)
\end{cases}
\end{equation}
\end{widetext}
The fiducial state (\ref{Dstate}) can be thus written as
\begin{equation}\label{fisher26}
\rho_0=\sum_{\alpha\alpha'\beta\beta'}c_{\alpha\beta}\,\overline{c}_{\alpha'\beta'}\,\tau^{(A)}_{\alpha\alpha'}\otimes\tau_{\beta\beta'}^{(B)}\;.
\end{equation}
where we do not explicitly write the superscripts $(k,\{j_e\})$ to simplify the notation. As discussed in appendix \ref{appendixA}, a direct computation of the expressions (\ref{fisher25}) yields
\begin{equation}\label{fisher27}
\mathcal K_{\substack{(aa')\\(bb')}}^{(AB)}=c_{a'b'}\,\overline{c}_{ab}-\sum_\beta c_{a'\beta}\,\overline{c}_{a\beta}\cdot\sum_\gamma c_{\gamma b'}\,\overline{c}_{\gamma b}\;,
\end{equation}
\begin{equation}\label{fisher29}
\begin{split}
\mathcal K_{\substack{(aa')\\(bb')}}^{(A)}&=\frac{1}{2}\Bigl(\delta_{a'b}\sum_\beta c_{b'\beta}\,\overline{c}_{a\beta}+\delta_{ab'}\sum_\beta c_{b\beta}\,\overline{c}_{a'\beta}\Bigr)-\\
&\quad - \sum_\beta c_{a'\beta}\,\overline{c}_{a\beta}\cdot\sum_\delta c_{b'\delta}\,\overline{c}_{b\delta}\;,
\end{split}
\end{equation}
\begin{equation}\label{fisher30}
\mathcal K_{\substack{[aa']\\ [bb']}}^{(A)}= \frac{1}{2}\Bigl(\delta_{a'b}\sum_\beta c_{b'\beta}\,\overline{c}_{a\beta}-\delta_{ab'}\sum_\beta c_{b\beta}\,\overline{c}_{a'\beta}\Bigr)\;.
\end{equation}
Similar results hold for $\mathcal K^{(BA)}$ and the $\mathcal K^{(B)}$ blocks of the symmetric and antisymmetric part, respectively.\\ \\Finally, omitting for the moment the constant factor in front of the trace in Eq. (\ref{fisher21}), the entanglement monotone $\mathcal E$ is given by (cfr. Eqs. (\ref{KtK},\ref{entmono})):
\begin{equation}\label{fisher32}
\mathcal E=\sum_{aa'cc'}\Bigl(c_{a'c'}\,\overline{c}_{ac}-\sum_\beta c_{a'\beta}\,\overline{c}_{a\beta}\sum_\gamma c_{\gamma c'}\,\overline{c}_{\gamma c}\Bigr)^2\;.
\end{equation}
Here we see the advantage of choosing the standard basis. Indeed, in this basis, our measure of entanglement $\mathcal E$, as well as all the block-coefficient matrices of the tensor $\mathcal K$, are written directly  in terms of the coefficients $c$ of the fiducial state.
\subsection{Some special  cases}\label{VE}
Let us finally check that the blocks of the tensor $\mathcal K$ actually encode the information about separability or entanglement by considering some explicit choice of the fiducial state $\rho_0$. In particular we have to check that the off-diagonal blocks, and hence the entanglement monotone $\mathcal E$, vanish when $\rho_0$ is separable, while the symplectic part vanishes when $\rho_0$ is maximally entangled. So if $\rho_0$ is separable, the coefficients $c_{\alpha\beta}$ factorize as $\lambda_\alpha\lambda_\beta$ and the fiducial state (\ref{fisher26}) can be written as:
\begin{equation}\label{fisher33}
\rho_0=\sum_{\alpha,\alpha'=1}^{N_A}\lambda_\alpha\bar\lambda_{\alpha'}\tau_{\alpha\alpha'}^{(A)}\otimes \sum_{\beta,\beta'=1}^{N_B}\lambda_\beta\bar\lambda_{\beta'}\tau_{\beta\beta'}^{(B)}\equiv\rho_0^{(A)}\otimes\rho_0^{(B)}\;.
\end{equation}
Therefore, when this is the case, the matrix elements of the off-diagonal blocks (\ref{fisher27}) are given by
\begin{equation}\label{fisher34}
\begin{split}
\mathcal K_{\substack{(aa')\\(bb')}}^{(AB)}&=\lambda_{a'}\lambda_{b'}\bar\lambda_a\bar\lambda_b-\lambda_{a'}\bar\lambda_a\Bigl(\sum_\beta\lambda_\beta\bar\lambda_\beta\Bigr)\cdot\lambda_{b'}\bar\lambda_b\Bigl(\sum_\gamma\lambda_\gamma\bar\lambda_\gamma\Bigr)\\
&=\lambda_{a'}\lambda_{b'}\bar\lambda_a\bar\lambda_b-\lambda_{a'}\bar\lambda_a\lambda_{b'}\bar\lambda_b =0
\end{split}
\end{equation}
where we have used the normalization condition $\sum_\beta|\lambda_\beta|^2=1$. Thus, being $\mathcal K^{(AB)}=0$ in the separable case, also the entanglement measure $\mathcal E$ defined in (\ref{fisher21}) obviously vanishes as it can be directly checked from Eq. (\ref{fisher32}).\\On the other hand, if $\rho_0$ is maximally entangled, then the coefficients $c_{\alpha\beta}$ are given by $\delta_{\alpha\beta}/\sqrt{N_<}$ with $N_<=\min{(N_A,N_B)}$ and the fiducial state can be written as:
\begin{equation}\label{fisher35}
\rho_0=\frac{1}{N_<}\sum_{\alpha,\alpha'=1}^{N_<}\tau_{\alpha\alpha'}^{(A)}\otimes\tau_{\alpha\alpha'}^{(B)}\;.
\end{equation}
From the expression (\ref{fisher30}) we then see that
\begin{align}\label{fisher36}
\mathcal K_{\substack{[aa']\\ [bb']}}^{(A)}&=\frac{1}{2N_<}\Bigl(\delta_{a'b}\sum_\beta\delta_{b'\beta}\delta_{a\beta}-\delta_{ab'}\sum_\beta\delta_{b\beta}\delta_{a'\beta}\Bigr)\\ \nonumber
&=\frac{1}{2N_<}\Bigl(\delta_{a'b}\delta_{b'a}-\delta_{ab'}\delta_{ba'}\Bigr)=0\;,
\end{align}
i.e., as expected, the symplectic part vanishes in the maximally entangled case. Moreover, in this case the entanglement measure (\ref{fisher21}) is given by
\begin{align}\label{fisher37}
\begin{split}
&\text{Tr}\left(\mathcal K^{(AB)\,T}\mathcal K^{(AB)}\right)=\\ 
&=\sum_{aa'cc'}\biggl(\frac{1}{N_<}\delta_{a'c'}\delta_{ac}-\frac{1}{N_<^2}\sum_\beta\delta_{a'\beta}\delta_{a\beta}\sum_\gamma\delta_{\gamma c'}\delta_{\gamma c}\biggr)^2\\ 
&=\sum_{aa'cc'}\biggl(\frac{1}{N_<}\delta_{a'c'}\delta_{ac}-\frac{1}{N_<^2}\delta_{aa'}\delta_{cc'}\biggr)^2
\end{split}
\end{align}
then explicitly, we have\footnote{The explicit example of a simple spin network graph with all spins fixed to $\frac{1}{2}$ is discussed in appendix \ref{appendixB}.}
\begin{align}\label{fisher37b}
&\text{Tr}\left(\mathcal K^{(AB)\,T}\mathcal K^{(AB)}\right)=\\ \nonumber
&=\sum_{aa'cc'}\biggl(\frac{1}{N_<^2}(\delta_{a'c'}\delta_{ac})^2+\frac{1}{N_<^4}(\delta_{a'a'}\delta_{cc'})^2+\\ \nonumber
&\quad -\frac{2}{N_<^3}\delta_{a'c'}\delta_{ac}\delta_{aa'}\delta_{cc'}\biggr)=1+\frac{1}{N_<^2}-\frac{2}{N_<^2}=\frac{N_<^2-1}{N_<^2}\;,
\end{align}
i.e., restoring the constant factor in Eq. (\ref{fisher32})
\begin{equation}\label{fisher38}
\mathcal E=\frac{N_<^2}{4(N_<^2-1)}\text{Tr}\left(\mathcal K^{(AB)\,T}\mathcal K^{(AB)}\right)=\frac{1}{4}\;,
\end{equation}
and the distance with respect to the separable state (\ref{fisher22}) takes the following value:
\begin{equation}
\text{Tr}\left(R^\dagger R\right)=\frac{1}{N_<^4}\text{Tr}\left(\mathcal K^{(AB)\,T}\mathcal K^{(AB)}\right)=\frac{N_<^2-1}{N_<^6}\;.
\end{equation}
Finally, let us consider the intermediate case of a generic entangled fiducial state, that is
\begin{equation}\label{fisher39}
c_{\alpha\beta}=f(\alpha)\delta_{\alpha\beta}\;,
\end{equation}
where $f(\alpha)$ is a complex functions satisfying the normalization condition $\sum_\alpha|f(\alpha)|^2=1$.

The fiducial state (\ref{fisher26}) can be thus written as
\begin{equation}\label{fisher41}
\rho_0=\sum_{\alpha,\alpha'}f(\alpha)\overline{f}(\alpha')\,\tau_{\alpha\alpha'}^{(A)}\otimes\tau_{\alpha\alpha'}^{(B)}\;;
\end{equation}
hence we have
\begin{widetext}
\begin{align}\label{fisher42}
\begin{split}
&\text{Tr}\left(\mathcal K^{(AB)\,T}\mathcal K^{(AB)}\right)=\sum_{aa'cc'}\Bigl(f(a')\overline{f}(a)\delta_{a'c'}\delta_{ac}-\sum_\beta f(a')\overline{f}(a)\delta_{a'\beta}\delta_{a\beta}\sum_\gamma f(c')\overline{f}(c)\delta_{\gamma c'}\delta_{\gamma c}\Bigr)^2\\ 
&=\sum_{aa'cc'}\Bigl(f(a')\overline{f}(a)\delta_{a'c'}\delta_{ac}-f(a')\overline{f}(a)\delta_{a'a}f(c')\overline{f}(c)\delta_{cc'}\Bigr)^2\\ 
&=\sum_a\overline{f}(a)^2\sum_{a'}f(a')^2+ \sum_{aa'cc'}f(a')^2\overline{f}(a)^2\delta_{aa'}\delta_{aa'}f(c')^2\overline{f}(c)^2\delta_{cc'}\delta_{cc'}-2\sum_{aa'cc'}f(a')^2\overline{f}(a)^2f(c')\overline{f}(c)\delta_{aa'}\delta_{a'c'}\delta_{ac}\delta_{aa'}\delta_{cc'}\\
&=\sum_a\overline{f}(a)^2\sum_{a'}f(a')^2-\sum_{ac}|f(a)|^4|f(c)|^2\delta_{ac} =\sum_a\overline{f}(a)^2\sum_{a'}f(a')^2-\sum_{a}|f(a)|^6\;.
\end{split}
\end{align}
\end{widetext}

Therefore, if the functions $f$ are real\footnote{This is actually the case of a Schmidt decomposition.}, Eq. (\ref{fisher42}) reduces to:
\begin{equation}\label{fisher43}
\text{Tr}\left(\mathcal K^{(AB)\,T}\mathcal K^{(AB)}\right)=1-\sum_{a}|f(a)|^6\;.
\end{equation}
In particular, we see that when $f(a)=1/\sqrt{N_<}\,,\forall a=1,\dots,N_<$, we recover the result (\ref{fisher37}) for the maximally entangled case.\\ \\To sum up, we collect the above results in the following table:
\begin{widetext}
\begin{center}
\begin{tabular}{|c | c | c |}
\hline
& &\\
$\mathbf{\rho_0}$ & $\mathbf{c_{\alpha\beta}}$ & $\mathbf{\text{Tr}\left(\mathcal K^{(AB)\,T}\mathcal K^{(AB)}\right)}$\\
\hline
\hline
& & \\
separable & $\lambda_\alpha\lambda_\beta$ & 0\\
& & \\
maximally \quad & &\\
entangled \quad & $\frac{\delta_{\alpha\beta}}{\sqrt{N_<}}$ & $1-\frac{1}{N_<^2}$\\
& & \\
 & \quad$f(\alpha)\delta_{\alpha\beta}\;,\; f(\alpha)\in\mathbb C$ \quad& \quad$\sum_\alpha\overline{f}(\alpha)^2\sum_{\alpha'}f(\alpha')^2-\sum_{\alpha}|f(\alpha)|^6$\\
entangled \quad & & \\
 &\quad $f(\alpha)\delta_{\alpha\beta}\;,\; f(\alpha)\in\mathbb R$ \quad & \quad $1-\sum_{\alpha}|f(\alpha)|^6$\\
 & & \\ 
\hline
\end{tabular}
\end{center}
\end{widetext}

\section{Conclusions and outlook}\label{out}

Motivated by the idea that, in the background independent framework of a quantum theory of gravity, entanglement is expected to play a key role in the reconstruction of spacetime geometry, this work is a preliminary investigation towards the possibility of using the formalism of Geometric Quantum Mechanics (GQM) to give a fully tensorial characterization of entanglement on spin network states. Given that such states also carry an intrinsic quantum geometric characterization in terms of the algebraic data labeling them, interpreted in the sense of simplicial geometry, an additional issue is to relate any geometric notion encoded in their entanglement properties with such simplicial geometry. Our analysis focused first on the simple case of a single link graph state for which we define a dictionary to construct a Riemannian metric tensor and a symplectic structure on the space of states. The manifold of (pure) quantum states was then stratified in terms of orbits of equally entangled states showing that the block-coefficient matrices of the corresponding pulled-back tensors fully encode the information about separability and entanglement. In particular, the off-diagonal blocks $C$ define an entanglement monotone $\mathcal E\propto\text{Tr}(C^TC)$, directly related to the geometric distance with respect to the separable state. Such a construction provides:

\begin{enumerate}
\item A formalism which fits well to a purely relational interpretation of the link as an elementary process describing the quantum correlations between its endpoints.
\item A quantitative characterization of graph connectivity by means of the entanglement monotone $\mathcal E$ which comes to be a measure of the existence of the process/link.
\item A connection between the GQM formalism and the geometric properties of the quantum states through entanglement. In the maximally entangled case, which for the single link corresponds to a gauge-invariant loop, the entanglement monotone is actually proportional to a power of the corresponding expectation value of the area operator. 
\end{enumerate}
As a second step, we applied the construction to the case of a spin network intertwiner state, dual to a fundamental volume of space. In this framework, we focussed on the Hilbert space of the single N-valent intertwiner and we regarded the whole system as a bipartite one, where each subsystem is $n$-level and the number of levels is determined by the degeneracy of the two virtual intertwiner spaces, resulting from the unfolding of the initial vertex. We then studied the resulting quantum correlations using our GQM formalism. The series of analytic results derived for the entanglement monotone $\mathcal E$ support its interpretation as a measure of spatial connectivity.

\begin{figure}[t]
\includegraphics[width=3.4 in]{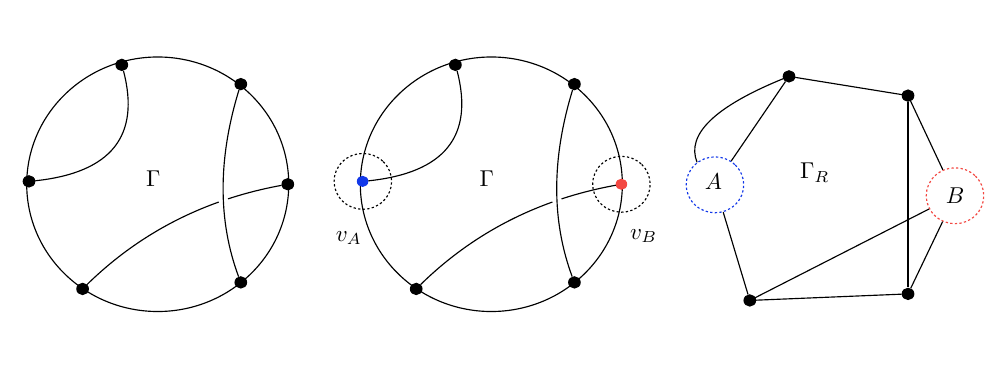}
\includegraphics[width=3.4 in]{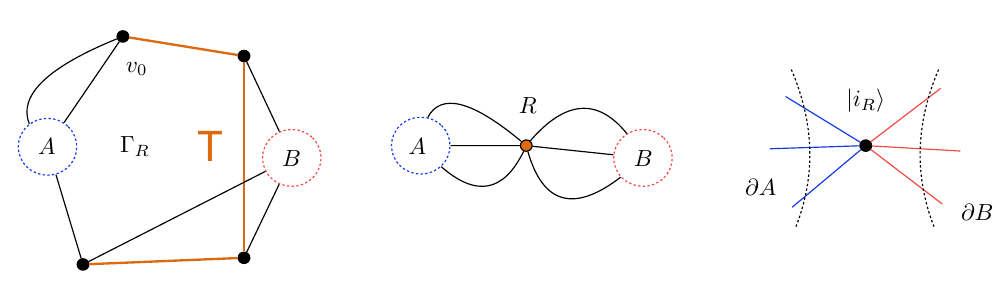}
\caption{Modeling of the bounded region with no curvature: single out two subregions (nodes $A$ and $B$) form the graph $\Gamma$ and consider the correlations among their boundaries ($\partial A$, $\partial B$) induced by the presence of the intermediate region $\Gamma_{R}$. By gauge fixing, $\Gamma_R$ reduces to a single node graph, intertwining the edges dual to the two boundaries.}\label{flower1}
\end{figure}

In fact, our interest in considering intertwiner states goes beyond their role of fundamental structural tensors attached to nodes. Indeed, more generally, within a reduction by gauge fixing scheme, intertwiner spaces provides a synthetic \emph{coarse grained} description for a generic closed region of quantum space with boundary and non-trivial internal degrees of freedom (see Fig. \ref{flower1}). Assuming bulk flatness, i.e. the absence of curvature degrees of freedom, a generic region of a spin network is effectively described by an intertwiner between the $n$ links puncturing the dual surface. From the point of view of the surface, a state of geometry of that region is described by a superposition of the possible $n$-valent intertwiners \cite{F17}. 

It has been proposed in \cite{F3} that a notion of distance between two regions of space should be derived in terms of the entanglement between the two regions A and B of the underlying spin network induced by the rest of the network. 
Our geometric approach along the same line, though still limited to the simplified case of a single $n$-valent intertwiner, however suggests that, in the flat bulk case,  the measure of entanglement does not depend on the simplicial bulk distance between the two non adjacent regions, which always trivialized due to the gauge invariance of the state, but only on the representations of the boundary states. 
\noindent
Overall, these results may be intended as the starting point of a program whose final goal aims to understand in full generality how the tensorial structures defined on the space of spin network states can be used to characterize their geometric features, alongside with (or as an alternative to) their simplicial geometric interpretation, and, in particular, how they can help in the reconstruction of the continuum (quantum) geometry of spacetime. We have seen that even at the simplest level of a single link we can already grasp some connections between entanglement and geometry. Obviously, the cases considered in this paper being so simple, we need to extend our construction to more general cases. Let us then close by sketching some possible future developments:
\begin{itemize}
\item The general setup we have used to compute entanglement properties of the bipartite system associated to two regions of a spin network state should now be extended, and the calculations generalized. This can be done in several directions, corresponding to the progressive removal of the various approximations we have imposed on our system, in this work. One is the introduction of curvature degrees of freedom, or of a proper coarse graining procedure to deal with them. Another is the inclusion of the sum over spin degrees of freedom in the calculations, i.e. allowing for a superposition of spin network states while keeping the underlying combinatorial structure fixed. More ambitiously, we need to learn to control the entanglement properties of superpositions over graph structures. For the latter goal, the GFT formalism may be the one with the greatest potential. In order to generalize our entanglement calculation to superpositions of spin network graphs, one possibility is to adopt the tensor model techniques already used for the calculation of the entanglement entropy of horizon states built out of spin networks in \cite{horizon}.
\item  We may focus on coherent states and exploit their interpretation as semiclassical states to study the classical limit of the metric tensor. Let us also notice that in this case we are selecting a particular family of states with their own parameter space. Moreover, since the parameter space of coherent states is isomorphic to the classical phase space of discrete geometries associated to spin network states, establishing a correspondence between the (entanglement) properties measured in terms of information geometry and the  same simplicial geometry should be rather direct. This should also enable us to exploit  the connection between the Fubini-Study and the Fisher-Rao metrics and the related tools of information geometry (see for example \cite{MMVV16} where the quantum metric is derived from relative entropy).
\item The analysis of entanglement with classical tensors has been extended also to the case of mixed states \cite{GQM8}. The case of Gibbs states, where expectation values of geometric observables such as area play the role of the parameters of the exponential family of maximally mixed states, would be interesting ion view of its application to the study of black holes. In particular, once we generalize our construction to include curvature degrees of freedom in the bulk of our spatial region, any subsequent coarse-graining of the same curvature degrees of freedom will result in mixed states. In other words, one could associate the mixed nature of quantum states being considered to the curvature degrees of freedom having been traced out. 
\item A further interesting aspect concerns the very interpretation of these tensors in those cases where the space of states is a tensor product of boundary states spaces of a process. The case of the single link, where the Hermitian tensor can be associated with an amplitude from an initial to a final spin state, may be generalized to a full (spin foam) path integral amplitude, meant as a process generating a region of space-time. In this case, the Fubini-Study metric would provide a metric for the space-time region. This setting has interesting formal analogies with the general boundary formalism \cite{F19,F20}.
\end{itemize}

We finally wish to comment about the novelty and relevance of the problem treated in the paper: the new element is the application of geometric quantum mechanics to the study of correlations on spin network states. The idea of using information theoretic tools in the context of quantum gravity is certainly not new, but the use of geometry in the context of quantum information has recently seen a great development (see \cite{MMVV16, tomo, HJIG} and references therein) and its application to quantum gravity is certainly new. These geometric techniques, in our opinion, will allow to import quantum information tools in quantum gravity in a more efficient and fruitful way, than it has been done so far. Tensorial structures defined on the space of quantum states of whatever dynamical system one is analyzing have the advantage of being basis independent and intrinsic. The results that we have obtained about the relation between spin networks connectivity and entanglement are certainly preliminary and for the moment limited to pure states. However, along with the quantum gravity perspective considered, the most compelling feature of this formalism resides in the possibility of generalizing the treatment to the case of mixed states, something precluded to any analysis built on entanglement entropy, as well as to the case of multipartite entanglement. Both aspects are expected to play a fundamental role once coarse graining and entanglement renormalisation schemes will be considered, for example in the study of the continuum limit leading to the emergence of semiclassical geometry.

\acknowledgments
F. M. thanks AEI Potsdam for hospitality and INFN for support. P.V.  acknowledges  support by COST (European Cooperation in Science  and  Technology)  in  the  framework  of  COST  Action  MP1405  QSPACE and hospitality at AEI Potsdam.

\newpage

\appendix

\section{Computation in the standard basis}\label{appendixA}
In this appendix we report the explicit computations of the block-coefficient matrices (\ref{fisher25}) of the Hermitian quantum Fisher tensor on the orbit generated from the fiducial state
\begin{equation}
\rho_0=\sum_{\alpha\alpha'\beta\beta'}c_{\alpha\beta}\,\overline{c}_{\alpha'\beta'}\,\tau^{(A)}_{\alpha\alpha'}\otimes\tau_{\beta\beta'}^{(B)}\;.
\end{equation}
Let us start with the off-diagonal blocks:
\begin{widetext}
\begin{equation}\label{offdiag}
\begin{split}
\mathcal K_{\substack{(aa')\\(bb')}}^{(AB)}&=\sum_{\substack{\alpha\alpha'\\ \beta\beta'}}c_{\alpha\beta}\,\overline{c}_{\alpha'\beta'}\,\text{Tr}\left(\tau_{\alpha\alpha'}^{(A)}\tau_{aa'}^{(A)}\otimes\tau_{\beta\beta'}^{(B)}\tau_{bb'}^{(B)}\right)-\sum_{\substack{\alpha\alpha'\\ \beta\beta'}}c_{\alpha\beta}\,\overline{c}_{\alpha'\beta'}\,\text{Tr}\left(\tau_{\alpha\alpha'}^{(A)}\tau_{aa'}^{(A)}\otimes\tau_{\beta\beta'}^{(B)}\right)\sum_{\substack{\gamma\gamma'\\ \delta\delta'}}c_{\gamma\delta}\,\overline{c}_{\gamma'\delta'}\,\text{Tr}\left(\tau_{\gamma\gamma'}^{(A)}\otimes\tau_{\delta\delta'}^{(B)}\tau_{bb'}^{(B)}\right)\\
&=\sum_{\alpha\alpha'\beta\beta'}c_{\alpha\beta}\,\overline{c}_{\alpha'\beta'}\,\text{Tr}_A\left(\tau_{\alpha\alpha'}^{(A)}\tau_{aa'}^{(A)}\right)\text{Tr}_B\left(\tau_{\beta\beta'}^{(B)}\tau_{bb'}^{(B)}\right)+\\
&-\sum_{\alpha\alpha'\beta\beta'}c_{\alpha\beta}\,\overline{c}_{\alpha'\beta'}\,\text{Tr}_A\left(\tau_{\alpha\alpha'}^{(A)}\tau_{aa'}^{(A)}\right)\text{Tr}_B\left(\tau_{\beta\beta'}^{(B)}\right)\cdot\sum_{\gamma\gamma'\delta\delta'}c_{\gamma\delta}\,\overline{c}_{\gamma'\delta'}\,\text{Tr}_A\left(\tau_{\gamma\gamma'}^{(A)}\right)\text{Tr}_B\left(\tau_{\delta\delta'}^{(B)}\tau_{bb'}^{(B)}\right)\\
&=\sum_{\alpha\alpha'\beta\beta'}c_{\alpha\beta}\,\overline{c}_{\alpha'\beta'}\,\delta_{\alpha a'}\delta_{\alpha' a}\delta_{\beta b'}\delta_{\beta' b}-\sum_{\alpha\alpha'\beta\beta'}c_{\alpha\beta}\,\overline{c}_{\alpha'\beta'}\,\delta_{\alpha a'}\delta_{\alpha'a}\delta_{\beta\beta'} \sum_{\gamma\gamma'\delta\delta'}c_{\gamma\delta}\,\overline{c}_{\gamma'\delta'}\,\delta_{\gamma\gamma'}\delta_{\delta b'}\delta_{\delta'b}\\
&=c_{a'b'}\,\overline{c}_{ab}-\sum_\beta c_{a'\beta}\,\overline{c}_{a\beta}\cdot\sum_\gamma c_{\gamma b'}\,\overline{c}_{\gamma b}
\end{split}
\end{equation}
\end{widetext}

where in the third equality we used the relations $\text{Tr}_A\bigl(\tau_{\alpha\alpha'}^{(A)}\tau_{aa'}^{(A)}\bigr)=\delta_{\alpha a'}\delta_{\alpha'a}$ and $\text{Tr}_A\bigl(\tau_{\gamma\gamma'}^{(A)}\bigr)=\delta_{\gamma\gamma'}$ (and similarly for the $\tau^{(B)}$'s) which can be easily deduced from Eq. (\ref{fisher23}). A similar result holds for the block $\mathcal K^{(BA)}$.\\As regards the other blocks, by using the commutation and anti-commutation relations
\begin{equation}\label{fisher28}
\bigl[\tau_{aa'}^{(X)},\tau_{bb'}^{(X)}\bigr]_\pm=\delta_{a'b}\tau_{ab'}^{(X)}\pm\delta_{ab'}\tau_{a'b}^{(X)}\,,\quad X=A,B
\end{equation}
we get
\begin{widetext}
\begin{align}\label{fisher29}
\mathcal K_{\substack{(aa')\\(bb')}}^{(A)}&=\frac{1}{2}\sum_{\alpha\alpha'\beta\beta'}c_{\alpha\beta}\,\overline{c}_{\alpha'\beta'}\text{Tr}_A\left(\tau_{\alpha\alpha'}^{(A)}\bigl[\tau_{aa'}^{(A)},\tau_{bb'}^{(A)}\bigr]_+\right)\text{Tr}_B\left(\tau_{\beta\beta'}^{(B)}\right)-\sum_{\alpha\alpha'\beta\beta'}c_{\alpha\beta}\,\overline{c}_{\alpha'\beta'}\text{Tr}_A\left(\tau_{\alpha\alpha'}^{(A)}\tau_{aa'}^{(A)}\right)\text{Tr}_B\left(\tau_{\beta\beta'}^{(B)}\right)\cdot \\ \nonumber
&\cdot\sum_{\gamma\gamma'\delta\delta'}c_{\gamma\delta}\,\overline{c}_{\gamma'\delta'}\text{Tr}_A\left(\tau_{\gamma\gamma'}^{(A)}\tau_{bb'}^{(A)}\right)\text{Tr}_B\left(\tau_{\delta\delta'}^{(B)}\right)\\ \nonumber
&=\frac{1}{2}\sum_{\alpha\alpha'\beta}c_{\alpha\beta}\,\overline{c}_{\alpha'\beta}\text{Tr}_A\left(\tau_{\alpha\alpha'}^{(A)}\bigl[\tau_{aa'}^{(A)},\tau_{bb'}^{(A)}\bigr]_+\right)-\sum_\beta c_{a'\beta}\,\overline{c}_{a\beta}\cdot\sum_\delta c_{b'\delta}\,\overline{c}_{b\delta}\\ \nonumber
&=\frac{1}{2}\sum_{\alpha\alpha'\beta}c_{\alpha\beta}\,\overline{c}_{\alpha'\beta}\left[\delta_{a'b}\text{Tr}_A\left(\tau_{\alpha\alpha'}^{(A)}\tau_{ab'}^{(A)}\right)+\delta_{ab'}\text{Tr}_A\left(\tau_{\alpha\alpha'}^{(A)}\tau_{a'b}^{(A)}\right)\right]-\sum_\beta c_{a'\beta}\,\overline{c}_{a\beta}\cdot\sum_\delta c_{b'\delta}\,\overline{c}_{b\delta}\\ \nonumber
&=\frac{1}{2}\sum_{\alpha\alpha'\beta}c_{\alpha\beta}\,\overline{c}_{\alpha'\beta}\left(\delta_{a'b}\delta_{\alpha b'}\delta_{\alpha'a}+\delta_{ab'}\delta_{\alpha b}\delta_{\alpha'a'}\right)-\sum_\beta c_{a'\beta}\,\overline{c}_{a\beta}\cdot\sum_\delta c_{b'\delta}\,\overline{c}_{b\delta}\\ \nonumber
&=\frac{1}{2}\Bigl(\delta_{a'b}\sum_\beta c_{b'\beta}\,\overline{c}_{a\beta}+\delta_{ab'}\sum_\beta c_{b\beta}\,\overline{c}_{a'\beta}\Bigr)-\sum_\beta c_{a'\beta}\,\overline{c}_{a\beta}\cdot\sum_\delta c_{b'\delta}\,\overline{c}_{b\delta}\;,
\end{align}

and
\begin{equation}\label{fisher30}
\begin{split}
\mathcal K_{\substack{[aa']\\ [bb']}}^{(A)}&=\frac{1}{2}\sum_{\alpha\alpha'\beta\beta'}c_{\alpha\beta}\,\overline{c}_{\alpha'\beta'}\text{Tr}_A\left(\tau_{\alpha\alpha'}^{(A)}\bigl[\tau_{aa'}^{(A)},\tau_{bb'}^{(A)}\bigr]_-\right)\text{Tr}_B\left(\tau_{\beta\beta'}^{(B)}\right)\\
&=\frac{1}{2}\sum_{\alpha\alpha'\beta}c_{\alpha\beta}\,\overline{c}_{\alpha'\beta}\text{Tr}_A\left(\tau_{\alpha\alpha'}^{(A)}\bigl[\tau_{aa'}^{(A)},\tau_{bb'}^{(A)}\bigr]_-\right)\\
&=\frac{1}{2}\sum_{\alpha\alpha'\beta}c_{\alpha\beta}\,\overline{c}_{\alpha'\beta}\left[\delta_{a'b}\text{Tr}_A\left(\tau_{\alpha\alpha'}^{(A)}\tau_{ab'}^{(A)}\right)-\delta_{ab'}\text{Tr}_A\left(\tau_{\alpha\alpha'}^{(A)}\tau_{a'b}^{(A)}\right)\right]\\
&=\frac{1}{2}\sum_{\alpha\alpha'\beta}c_{\alpha\beta}\,\overline{c}_{\alpha'\beta}\left(\delta_{a'b}\delta_{\alpha b'}\delta_{\alpha'a}-\delta_{ab'}\delta_{\alpha b}\delta_{\alpha'a'}\right)\\
&=\frac{1}{2}\Bigl(\delta_{a'b}\sum_\beta c_{b'\beta}\,\overline{c}_{a\beta}-\delta_{ab'}\sum_\beta c_{b\beta}\,\overline{c}_{a'\beta}\Bigr)\;.
\end{split}
\end{equation}
\end{widetext}

Similar results hold for the $\mathcal K^{(B)}$ blocks of the symmetric and antisymmetric part, respectively.\\Finally, let us consider the entanglement measure (\ref{fisher21}). First of all, let us compute the product $\mathcal K^{(AB)}\mathcal K^{(AB)\,T}$ whose matrix elements, according to Eq. (\ref{offdiag}), are given by
\begin{widetext}
\begin{equation}\label{KtK}
\begin{split}
\left(\mathcal K^{(AB)}\mathcal K^{(AB)\,T}\right)_{\substack{(aa')\\(bb')}}&=\sum_{cc'}\mathcal K_{\substack{(aa')\\(cc')}}^{(AB)}\mathcal K_{\substack{(cc')\\(bb')}}^{(AB)\,T}=\sum_{cc'}\mathcal K_{\substack{(aa')\\(cc')}}^{(AB)}\mathcal K_{\substack{(bb')\\(cc')}}^{(AB)}\\
&=\sum_{cc'}\Bigl[\Bigl(c_{a'c'}\,\overline{c}_{ac}-\sum_\beta c_{a'\beta}\,\overline{c}_{a\beta}\cdot\sum_\gamma c_{\gamma c'}\,\overline{c}_{\gamma c}\Bigr)\cdot\Bigl(c_{b'c'}\,\overline{c}_{bc}-\sum_\delta c_{b'\delta}\,\overline{c}_{b\delta}\cdot\sum_\zeta c_{\zeta c'}\,\overline{c}_{\zeta c}\Bigr)\Bigr]\\
&=\sum_{cc'}\Bigl(c_{a'c'}\,\overline{c}_{ac}c_{b'c'}\,\overline{c}_{bc}+\sum_\beta c_{a'\beta}\,\overline{c}_{a\beta}\sum_\gamma c_{\gamma c'}\,\overline{c}_{\gamma c}\sum_\delta c_{b'\delta}\,\overline{c}_{b\delta}\,\cdot\\
&\qquad\quad\cdot\sum_\zeta c_{\zeta c'}\,\overline{c}_{\zeta c}-c_{b'c'}\,\overline{c}_{bc}\sum_\beta c_{a'\beta}\,\overline{c}_{a\beta}\sum_\gamma c_{\gamma c'}\,\overline{c}_{\gamma c} -c_{a'c'}\,\overline{c}_{ac}\sum_\delta c_{b'\delta}\,\overline{c}_{b\delta}\sum_\zeta c_{\zeta c'}\,\overline{c}_{\zeta c}\Bigr)\;.
\end{split}
\end{equation}
\end{widetext}

Now taking the trace of Eq.(\ref{KtK}) amounts to set $a=b, a'=b'$ and to sum over $a,a'$. Therefore, omitting for the moment the constant factor in front of the trace in Eq. (\ref{fisher21}), we have:
\begin{widetext}

\begin{equation}\label{entmono}
\begin{split}
\mathcal E&=\sum_{aa'cc'}\Bigl[c_{a'c'}^2\,\overline{c}_{ac}^2+\Bigl(\sum_\gamma c_{\gamma c'}\,\overline{c}_{\gamma c}\Bigr)^2\Bigl(\sum_\delta c_{a'\delta}\,\overline{c}_{a\delta}\Bigr)^2-2c_{a'c'}\,\overline{c}_{ac}\sum_\beta c_{a'\beta}\,\overline{c}_{a\beta}\sum_\gamma c_{\gamma c'}\,\overline{c}_{\gamma c}\Bigr]\\
&=\sum_{aa'cc'}\Bigl(c_{a'c'}\,\overline{c}_{ac}-\sum_\beta c_{a'\beta}\,\overline{c}_{a\beta}\sum_\gamma c_{\gamma c'}\,\overline{c}_{\gamma c}\Bigr)^2\;.
\end{split}
\end{equation}
\end{widetext}

\section{Spin-$\frac{1}{2}$ graph and large $n$ correlations}\label{appendixB}
%\textcolor{blue}{As an explicit example for ...}
Let us consider the case of a spin network graph $\Gamma$ whose edges are all labeled by spins fixed at the fundamental representation, i.e., $j_e=\frac{1}{2}\;\forall e\in\partial A\cup\partial B\equiv\partial R$. The specific structure of the starting graph is not relevant for our analysis and the only assumption we make is that the gauge reduction procedure leads to a single intertwiner graph $\Gamma_R$ between the $2n$ ($n\in\mathbb N$) $SU(2)$-representations defining the boundary $\partial R$ with no loops carrying curvature excitations. The number of boundary edges must be necessarily even since there does not exist any intertwiner between an odd number of $\frac{1}{2}$-spin representations.

With such a choice of spin labels, we may unfold the single $2n$-valent intertwiner into two $(n+1)$-valent vertices coupling the two boundary sets of $n$ spin-$\frac{1}{2}$ edges with a virtual link labeled by a fixed spin $k$. As long as $k \leq n$, the dimension of the corresponding degeneracy spaces $\mathcal D_k^A$ and $\mathcal D_k^B$ can be then expressed in terms of binomial coefficients as \cite{lt1}:
\begin{align}\label{degfactor}
N=N^k_A=N^k_B&=d^{(n)}_k=\binom{2n}{n+k}-\binom{2n}{n+k+1}\\ \nonumber
&=\frac{2k+1}{n+k+1}\binom{2n}{n+k}\;.
\end{align}
Therefore, with this separation of boundary degrees of freedom, the gauge-reduced unfolded intertwiner state can be actually described as a couple of entangled $N$-level systems with the same number of levels given by (\ref{degfactor}).
Let us analize the large $n$ behaviour of the quantum correlations between the two subsystems as a function of $k$. We focus on a fiducial state $\ket{0}\equiv\ket{\psi_{\Gamma R}}$ with all coefficients $c_{\alpha_k\beta_k}^{(k,\{j_e=\frac{1}{2}\})}$ equal to $\frac{1}{\sqrt{N}}$ whose corresponding pure state density matrix is given by
\begin{equation}
\rho_0=\frac{1}{N}\sum_{\alpha_k,\alpha'_k=1}^N\tau_{\alpha_k\alpha'_k}^{(A)}\otimes\tau_{\alpha_k\alpha'_k}^{(B)}\;.
\end{equation}
In section V-E we found  that, for such a choice of the fiducial state, the entanglement measure constructed with the off-diagonal blocks of the pulled-back metric tensor on the orbit starting at $\rho_0$ is given by
\begin{equation}\label{monotone}
\text{Tr}\left(\mathcal K^{(AB)\,T}\mathcal K^{(AB)}\right)=1-\frac{1}{N^2}\;.
\end{equation}
For large $n$, say $n>>1$, the degeneracy factors (\ref{degfactor}) admit the following asymptotic expression \cite{etbe}
\begin{equation}\label{largen}
N=d^{(n)}_k\sim\frac{2^{2n+1}}{\sqrt{\pi n}}\frac{x}{(1+x)\sqrt{1-x^2}}\,e^{-n\varphi(x)}\;,
\end{equation}
with
\begin{equation}
\varphi(x)=(1+x)\log(1+x)+(1-x)\log(1-x), 
\end{equation}
for $x=\frac{k}{n}\in[0,1]$.
So now, when $k$ is much smaller than $n$ or equivalently $x\rightarrow\varepsilon$ with $\varepsilon<<1$, up to terms $o(\varepsilon^2)$ we have
\begin{equation}\label{ksmall}
N=d_k^{(n)}\sim\frac{2^{2n+1}}{\sqrt{\pi n}}\varepsilon+o(\varepsilon^2)\;.
\end{equation}
On the other hand, when $k$ becomes comparable with $n$ (that is $x\rightarrow1$), $\varphi(x)\sim2\log{2}$ and $d^{(n)}_{k=n}$ goes to infinity. Hence, we have:
\begin{equation}\label{limits}
N=d^{(n)}_k\underset{\text{large}\,n}{\sim}\begin{cases}
\frac{2^{2n+1}}{\sqrt{\pi n}}\varepsilon & \text{for}\;\;x\rightarrow\varepsilon\;\;(k<<n)\\
\infty & \text{for}\;\;x\rightarrow1\;\;(k\simeq n)
\end{cases}\;.
\end{equation}
The entanglement monotone (\ref{monotone}) then exhibits the following large $n$ behaviours
\begin{align}\label{largenmono}
\text{Tr}\left(\mathcal K^{(AB)\,T}\mathcal K^{(AB)}\right)&=1-\frac{1}{N^2}=1-\frac{1}{\bigl(d^{(n)}_k\bigr)^2}\\ \nonumber
&\sim\;\begin{cases}
1-\frac{\pi n}{(\varepsilon2^{2n+1})^2}& \text{for}\;\;k<<n\\
1 & \text{for}\;\;k\simeq n
\end{cases}\;.
\end{align}
Therefore, being (\ref{ksmall}) positive and greater than 1 (for large $n$ the numerator is greater than the denominator), the entanglement measure (\ref{largenmono}) reaches its maximum value when $k$ becomes comparable with $n$.

\section{Spin Networks in the Embedded Canonical Framework}\label{spin}
%We provide here a concise introduction to spin network states, focussing on their interpretation as elementary building blocks of quantum spacetime. 
%\subsection{Spin networks from canonical Loop Quantum Gravity}

In Loop Quantum Gravity, Einstein's theory of General Relativity (GR) is recast into the form of a \emph{gauge theory} with structure group $Spin(1,3)$, plus the additional gauge symmetries resulting from the space-time diffeomorphism invariance. 
A partial fixing of the $Spin(1,3)$ invariance leads to a phase space description of the classical theory in terms of connections $A$ of a principal $SU(2)$-bundle over spacelike hypersurfaces $\Sigma$, embedded in a spacetime manifold $\mathcal{M}$, and sections $E$ of the associated vector bundle over $\Sigma$, whose pull back  are Lie algebra valued pseudo two forms. 

The two forms $E$ encode the information about the 3d geometry on $\Sigma$, while the $A$ carries the information about the extrinsic curvature of $\Sigma$ in $\mathcal{M}$. 

The conjugated variables $(A,E)$, with standard Poisson brackets, define a (Yang--Mills like) phase space, which is then reduced by the imposition of the $SU(2)$ Gauss constraint, the spatial diffeomorphism constraint and the Hamiltonian constraint, respectively implementing the internal local gauge symmetry and the symmetry under diffeomorphisms.
 
 The Dirac quantization procedure, before the imposition of the diffeomorphism constraints, leads to Hilbert spaces $\mathcal{H}_\Gamma$ associated to graphs embedded in the canonical manifold.~\footnote{For a  comprehensive introduction to the LQG quantum geometry states we refer to the literature cited in the introduction, as well as to \cite{LQG6,LQG7,LQG8,LQG9,LQG10,LQG11,LQG12,LQG13,LQG14}.
} 
 
%\subsection{Quantum Geometry States over Embedded Spin Network Graphs}
Let $\Gamma\subset\Sigma$ be a graph, i.e., a finite and ordered collection of smooth oriented paths $\gamma_\ell\in\Sigma$ with $\ell=1,\dots,L$ meeting at most at their endpoints (such paths will be called the \textit{links} or the \textit{edges} of the graph, while the intersection points will be called \textit{nodes} or \textit{vertices}), and let $\psi: SU(2)^L\rightarrow\mathbb C$ be a (smooth) (cylindrical) function $\psi_\Gamma(h_1,\dots,h_L)$ of $L$ group elements. These group elements are interpreted as parallel transports $h_\ell(A)\equiv h_{\gamma_\ell}(A)$ of the connection $A$ along the links $\gamma_\ell$ of the graph $\Gamma$, embedded in the canonical hypersurface. 
The linear space of such cylindrical functionals w.r.t. a given graph $\Gamma$
can be turned into a Hilbert space by equipping it with the following scalar product %between two functionals constructed on the same graph
\begin{equation}\label{qg54}
\braket{\psi_{(\Gamma)}|\psi'_{(\Gamma)}}\equiv\int\prod_{\ell=1}^Ldh_\ell\,\overline{\psi(h_1,\dots,h_L)}\,\psi'(h_1,\dots,h_L)\;,
\end{equation}
where $dh_\ell$ are $L$ copies of the (left- and right-invariant) Haar measure of $SU(2)$. The inner product (\ref{qg54}) is invariant under $SU(2)$ gauge transformations acting as left or right multiplications on the arguments of the wave functions $\psi_\Gamma$, depending on whether the gauge transformation is associated to the starting or end point of the link to which each argument is referring to. This is a direct consequence of the invariance of the Haar measure. One then needs to construct a Hilbert space out of the space of all cylindrical functions for all graphs $\Gamma\subset\Sigma$: 
\begin{equation}\label{union}
\bigcup_{\Gamma\subset\Sigma}\mathcal H_\Gamma\;.
\end{equation}

To do this, we need to define a scalar product for cylindrical functions based on different graphs. Such a scalar product can be deduced from that on $\mathcal H_\Gamma$ as follows. The construction is based on the introduction of the so-called \textit{cylindrical equivalence relations} which reflect properties of the underlying continuum connection field, and it is therefore directly inspired by the continuum embedding of the graphs $\Gamma$, and thus on the origin of the quantum states $\psi_\Gamma$ as coming from the canonical quantization of a continuum field theory. The details of the construction are not so important for our purposes. Essentially,% we define equivalence classes of graphs that can be regarded as subgraphs of a bigger one, i.e.:
%\begin{equation}\label{qg59}
%[\Gamma]=\bigl\{\Gamma_1\sim\Gamma_2\quad\text{iff}\quad\exists\,\Gamma\subset\Sigma\;:\;\Gamma\supset\Gamma_1,\Gamma_2\bigr\}\;.
%\end{equation}
%and the inner product between states associated to different graphs can be defined to be like in (\ref{qg54}) for such bigger graph by defining
%\begin{equation}\label{qg60}
%\braket{\psi_{\Gamma_1}|\psi'_{\Gamma_2}}\equiv\braket{\psi_{\Gamma}|\psi'_{\Gamma}}\;,
%\end{equation}
%where $\Gamma=\Gamma_1\cup\Gamma_2$ and the functions $\psi_{\Gamma_1},\psi'_{\Gamma_2}$ are trivially extended on $\Gamma$ by setting them constant over the links which do not belong to $\Gamma_1,\Gamma_2$, respectively\footnote{Obviously, this new product reduces to the previous one if $\Gamma_1$ and $\Gamma_2$ coincide.}. The (unconstrained) kinematical Hilbert space will be therefore given by:
%\begin{equation}\label{qg58}
%\mathcal H_{kin}=\frac{\bigcup_{\Gamma\subset\Sigma}\mathcal H_\Gamma}{\sim}\;.
%\end{equation}
%In the spin representation, the equivalence condition $\sim$ which allows us to define the scalar product (\ref{qg60}) amounts to take all spins zero on the ``virtual'' links of the extended graph which do not belong to the starting one \cite{LQG14}. This implies that the Hilbert space (\ref{qg58}) can 
 the (unconstrained) kinematical Hilbert space of the theory can be casted as a direct sum of single graph-based Hilbert spaces
\begin{equation}\label{lqgkin}
\mathcal H_{kin}=\frac{\bigcup_{\Gamma\subset\Sigma}\mathcal H_\Gamma}{\sim}=\bigoplus_{\Gamma\subset\Sigma}\tilde{\mathcal H}_\Gamma\;,
\end{equation}
where the individual graph-based Hilbert spaces $\tilde{\mathcal H}_\Gamma$ correspond to $\mathcal H_\Gamma$ without zero modes, i.e., where the spins $j_\ell$ never take the value zero.
This space %equipped with the above inner product 
can then be understood as a Hilbert space over ``generalized'' connections on $\Sigma$ with the so-called Ashtekar-Lewandowski measure $d\mu_{AL}$ \cite{LQG32,LQG33,LQG34},
\begin{equation}\label{qg61}
\mathcal{H}_{kin}\cong L^2(A,d\mu_{AL})\;.
\end{equation}
In our analysis, we mainly focus on the spaces associated to single graphs $\Gamma$, thus with $\mathcal{H}_\Gamma$, and the cylindrical equivalence conditions will not play much of a role.

From such fundamental (unconstrained) kinematical Hilbert space $\mathcal H_{kin}$, the physical space $\mathcal H_{phys}$ is derived from a series of reduction processes under the imposition of the Gauss and diffeomorphism constraints. 

\subsection{Gauge-invariant states and spin network basis}
In this work we content ourselves with the kinematical structure of LQG, encoded in $\mathcal{H}^0_{kin}$, obtained from $\mathcal{H}_{kin}$ after the imposition of the Gauss constraint only, which is common to a large extent also to the group field theory formalism, as we will discuss.

The solutions of the quantum Gauss constraint form the Hilbert space $\mathcal H_{kin}^0$ of $SU(2)$-gauge invariant states, i.e.:
\begin{equation}\label{qg64}
\mathcal H_{kin}^0\equiv \text{Inv}_{SU(2)}\bigl[\mathcal H_{kin}\bigr]\; ,
\end{equation}
which can be defined by the same construction outlined in the previous section, but starting from gauge invariant spaces $\mathcal{H}^0_\Gamma$ associated to all possible graphs $\Gamma$.
From the transformation of continuum parallel transports under $SU(2)$ transformations, it follows that a gauge transformation acts only on the nodes of the graph. Therefore, the gauge-invariance requirement for cylindrical functions translates into the requirement of invariance under the action of the group at the nodes, i.e.:
\begin{align}\label{qg66}
&\psi^0_\Gamma(h_1,\dots,h_L)=\\ \nonumber
&=\psi^0_\Gamma \bigl(g(\gamma_1(0))h_1 g^{-1}(\gamma_1(1)),\dots,g(\gamma_L(0))h_L g^{-1}(\gamma_L(1))\bigr)\;,
\end{align}
where $\gamma_i(0)$ (resp. $\gamma_i(1)$) indicates the starting (resp. end) point of the link $i$ of the graph $\Gamma$. 
The above invariance can be implemented by group averaging
\begin{equation}\label{qg67}
%\psi^0_\Gamma(h_1,\dots,h_L)=
\int\prod_{v=1}^Vdg_v\,f\bigl(g(\gamma_1^s)h_1 g^{-1}(\gamma_1^e),\dots,g(\gamma_L^s)h_L g^{-1}(\gamma_L^e)\bigr)\;,
\end{equation}
where $V$ is the number of nodes (vertices) of the graph $\Gamma$ and we write $\gamma^{s} \equiv \gamma(0)$ (resp. $\gamma^{e} \equiv \gamma(1)$) for short notation. In the spin representation of each function, this corresponds to inserting on each node $v$ of the graph the following projector,
\begin{equation}\label{qg68}
\mathcal I_v=\int dg\,\prod_{\ell\in v}D^{(j_\ell)}(g)\;.
\end{equation}
with
\begin{equation}\label{qg69}
\prod_{\ell\in v}D^{(j_\ell)}_{m_\ell n_\ell}(g)\,\in\,\bigotimes_{\ell\in v}\mathcal H^{(j_\ell)}=\bigotimes_{\ell\in v}(\mathcal V^{(j_\ell)}\otimes {\mathcal V}^{(j_\ell)*})
\end{equation}
$\mathcal V^{(j_\ell)}$ denoting the $SU(2)$ irreducible spin-$j_\ell$ representation spaces. Therefore, by using the decomposition of the tensor product $\bigotimes_{\ell\in v}\mathcal H^{(j_\ell)}$ into irreducible representations
\begin{equation}\label{qg70}
\bigotimes_{\ell\in v}\mathcal H^{(j_\ell)}=\bigoplus_i\mathcal H^{(J_i)}\,,
\end{equation}
we find that $\mathcal I_v$ projects onto the gauge invariant part of $\bigotimes_{\ell\in v}\mathcal H^{(j_\ell)}$, namely the singlet space $\mathcal H^{(0)}$:
\begin{equation}\label{qg71}
\mathcal I_v\;:\;\bigotimes_{\ell\in v}\mathcal H^{(j_\ell)}\longrightarrow\mathcal H^{(0)}\,.
\end{equation}
Being $\mathcal I_v$ a projector, it can be decomposed in terms of a basis $\{i_\alpha\}$ of $\mathcal H^{(0)}$ and its dual as
\begin{equation}\label{qg72}
\mathcal I_v=\sum_{\alpha=1}^{\text{dim}\,\mathcal H^{(0)}}i_\alpha i_\alpha^*\;\in\;\mathcal H^{(0)}\otimes\mathcal H^{(0)*}\;,
\end{equation}
from which, together with the decomposition of $\bigotimes_{\ell\in v}\mathcal H^{(j_\ell)}=(\bigotimes_{\ell\,in}\mathcal H^{(j_\ell)*})\otimes(\bigotimes_{\ell\,out}\mathcal H^{(j_\ell)})$ between ingoing and outgoing links of the vertex $v$, it follows that $\mathcal I_v$ is the invariant map between the representation spaces associated with the edges joined at the node $v$, i.e.:
\begin{equation}\label{qg73}
\mathcal I_v\;:\;\bigotimes_{\ell\,in}\mathcal H^{(j_\ell)}\longrightarrow\bigotimes_{\ell\,out}\mathcal H^{(j_\ell)}\;.
\end{equation}
Such invariants are called \textit{intertwiners}. Hence, if we have an $p$-valent node, the intertwiner is an element of the invariant subspace $\text{Inv}_{SU(2)}\bigl[\mathcal H^{(j_1)}\otimes\dots\otimes\mathcal H^{(j_p)}\bigr]$ of the tensor product space between the $p$ irreducible representations associated to the links joining that node. However, such a procedure is possible only if some conditions necessary to have an invariant subspace are satisfied. For instance, in the case of a 3-valent node, there exists an intertwiner space only if the spin numbers $j_1,j_2,j_3$ labelling the representations associated to the three links satisfy the Clebsch-Gordan condition:
\begin{equation}\label{qg74}
|j_1-j_2|\leq j_3 \leq j_1+j_2\;.
\end{equation}
For a $p$-valent node (with $p>3$) the space $\mathcal H^{(0)}$ can have a larger dimension and the construction consists of adding first two irreducible representations, then the third, and so on, thus giving rise to a decomposition in virtual 3-valent nodes in which virtual links are labelled by spins $k$ satisfying the condition (\ref{qg74}). 
%\begin{figure}[t!]
%\centering
%\includegraphics[scale=0.40]{intertwiner}
%\caption{\textit{Construction of an intertwiner for a 4-valent vertex.}}
%\label{inter}
%\end{figure}

Since the projector (\ref{qg72}) acts only on the nodes of the graph that labels the basis of $\mathcal H_{kin}$, we can write the result of the action of $\mathcal I_v$ on elements of $\mathcal H_{kin}$ as a linear combination of products of representation matrices $D^{(j_\ell)}_{m_\ell n_\ell}(h_\ell(A))$ contracted with intertwiners. This leads us to give the following\\

\noindent \textbf{Definition:} {\it A triplet $(\Gamma, \vec j, \vec i)$ representing a graph $\Gamma$ embedded in $\Sigma$ whose $L$ links are colored by the spins $\vec j=(j_1, \dots, j_L)$ and whose $V$ nodes are labelled by intertwiners $\vec i=(i_1,\dots,i_V)$ is called a \textbf{spin network} $S$ embedded in $\Sigma$ associated with the graph $\Gamma$. A \textbf{spin network state} $\ket{S}\equiv\ket{\Gamma;\vec j,\vec i}$ is  the cylindrical function over the spin network $S$ associated with the graph $\Gamma$ which can be written as
\begin{equation}\label{qg75}
\braket{A|\Gamma;\vec j,\vec i}=\psi_{\Gamma,\vec j,\vec i}[A]=\bigotimes_\ell D^{(j_\ell)}(h_\ell(A))\cdot\bigotimes_vi_v\;,
\end{equation}
where $D^{(j_\ell)}(h_\ell(A))$ are the spin irreducible representations of the holonomy along each link and $\cdot$ denotes the contraction with the intertwiners whose indices (hidden for simplicity) can be reconstructed from the connectivity of the graph.}\\

These states form a complete orthonormal basis for $\mathcal{H}_\Gamma^0$ \cite{LQG37}%Indeed, using the Peter-Weyl theorem according to which the Wigner matrices form an orthonormal basis of $L^2(SU(2))$, and the definition of the scalar product (\ref{qg60}), we have:
%\begin{equation}\label{qg76}
%\braket{\Gamma';\vec j\,',\vec i\,'|\Gamma;\vec j,\vec i}\equiv\braket{\psi_{\Gamma';\vec j\,',\vec i\,'}|\psi_{\Gamma;\vec j,\vec i}}=\delta_{\Gamma',\Gamma}\,\delta_{\vec j\,',\vec j}\,\delta_{\vec i\,',\vec i}\;.
%\end{equation}
%The $SU(2)$ constraint is implemented by choosing an intertwiner at each node as discussed before. Thus, 
, given by
\begin{eqnarray}\label{qg77}
\mathcal H^0_\Gamma&=&\bigoplus_{j_\ell}\bigl(\bigotimes_v Inv_{SU(2)}\bigl[\bigotimes_{\ell\in v}\mathcal H^{(j_\ell)}\bigr]\bigr)\\ \nonumber
&\cong& L^2\left(SU(2)^L/SU(2)^V\right)
\end{eqnarray}
for a fixed graph $\Gamma$ with $L$ links and $V$ nodes.
%Spin networks will be the main focus of our attention in this paper. 
%As we mentioned, s
\begin{figure}[t!]
\includegraphics[scale=.55]{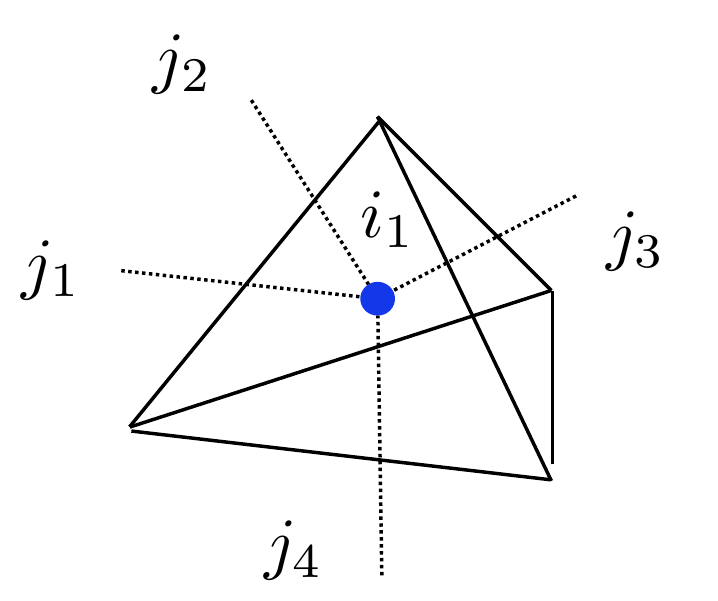}
\caption{Heuristic picture of a minimal chunk of quantum space: a tetrahedron dual to a single node of a spin network state. A discrete 3d quantum geometry is realised by a superposition of quantum tetrahedra, glued together by specific adjacency conditions.}
\label{CHUNK}
\end{figure}
Spin network states diagonalize geometric operators such as area and volume. 
In particular, %as only nodes contribute to the spectrum of the volume operator, the volume of a given region of a spin network is actually the sum of the $v$ volume contributions of the nodes inside the region. 
the face dual to each link $\ell$ has an area proportional to the spin label $j_\ell$, and each region around a node $v$ has a volume determined by the intertwiner $i_v$. This matches the identification of spin networks with states of quantum polyhedra dual to the nodes of the graph \cite{LQG6} and, more heuristically, traduces into a picture where $3$d space-like surfaces are represented by a collection of ``chunks'' (the polyhedra dual to the nodes, see Fig.~(\ref{CHUNK})) with quantized volume, which share surfaces whose area is determined by the spin of the dual link connecting them. Due to the embedding into a differentiable manifold, the algebraic data set colouring the LQG spin network graph provides a notion of quantum geometry which is at the same time discrete and relational \cite{LQG58}.
%However, if we get rid of the geometric characterisation provided by the embedded setting, how much of the space geometry could we extract from the structure of correlation of such states?

\section{Spin Networks as Quantum Many--Body States}\label{gft}
% symmetric tensor network states (hence still endowed with their pre-geometric character\footnote{For a general conceptual discussion on the meaning of ``pregeometry'', see for instance \cite{LQG61}.})

A different approach along the same line, consists in starting from \emph{abstract} spin network structures, with states defined independently of any embedding into a continuum manifold, hence with no reference to background notions of space, time or geometry \cite{LQG60}. 
This is the way spin networks appear in the group field theory formalism.\footnote{The possibility of defining spin network states in a more abstract, combinatorial way has been considered also within the canonical LQG approach \cite{absn1,absn2}.}

%In the lack of a background, adjacent regions of a spin network will not necessarily correspond to close regions of space. More generally, being the fundamental structure of space a quantum superposition of abstract non-embedded entities each of which having a different connectivity (i.e., a different graph structure), what is local in one term of the superposition will in general not be local in others \cite{LQG56,LQG57}. No metric structure allows to define a notion of distance and there is no absolute position at all. However, a given region of a spin network can still be localized with respect to other parts of the graph. This picture suggests that notions as  ``close'' and ``far'' should be reconsidered in terms of quantum correlations between subregions of the spin network graph as well as relations between spin network states in the Hilbert space.
%In this sense, quantum information geometry (geometric quantum mechanics?) should provide new crucial tools to investigate how geometry is encoded into the entanglement structure of the states. This is the main focus of our work.
%
%Before proceeding with our analysis of entanglement in spin network states, we need to provide further details on the notion of generalised spin network states with their associated Hilbert spaces.
%
%
%\subsection{Gluing Nodes via Links }

Along this line, in particular, spin networks can be  reformulated as quantum ``many-body'' systems, where  closed graphs result from a precise gluing prescription among individual single vertex open graphs (tensors) dual to fundamental volumes of space. 

\begin{figure}[t!]
\centering
\includegraphics[scale=0.24]{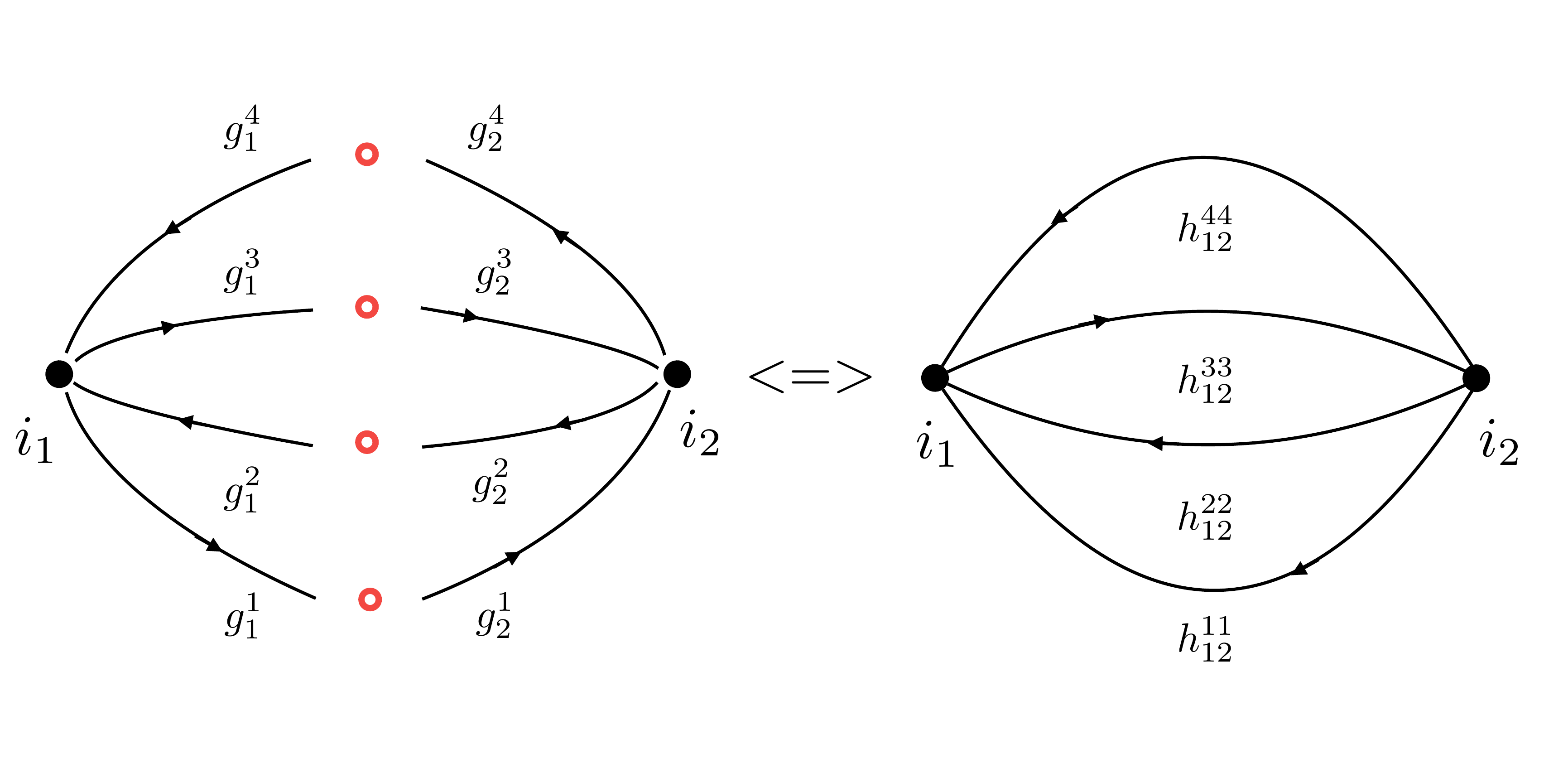}
\caption{Gluing of open spin network vertices to form a spin network closed graph.}
\label{glu}
\end{figure}

Let us consider a closed graph $\Gamma$ with $V$ $d$-valent vertices labelled by the index $i=1,\dots,V$ and denote the set of its edges by
\begin{equation}
L(\Gamma)=(\{1,\dots,V\}\times\{1,\dots,d\})^2
\end{equation}
such that
\begin{equation}
[(ia)(ia)]\notin L(\Gamma)\qquad,\qquad [(ia)(jb)]\in L(\Gamma)
\end{equation}
where the last condition specifies the connectivity of the graph telling us the existence of a directed edge connecting the $a$-th link at the $i$-th node to the $b$-th link at the $j$-th node, with source $i$ and target $j$. A generic cylindrical function based on the graph $\Gamma$ will be a function of the group elements $h_{ij}^{ab}\in\mathbb G$ ($\mathbb G\equiv SU(2)$ in LQG) assigned to each link $\ell:=[(ia)(jb)]\in L(\Gamma)$\footnote{Since in what follows it is important to distinguish different nodes and the links joining them, we admit a certain excessive complexity of notation using more indices to label links ($ab$) and their source and target nodes ($ij$).}
\begin{equation}
\psi_\Gamma(h_{12}^{11},h_{13}^{21},\dots)=\psi_\Gamma(\{h_{ij}^{ab}\})\;\in\;\mathcal H_{\Gamma}\cong L^{2}(\mathbb G^L/\mathbb G^V)\;,
\end{equation}
with $h_{ij}=h_{ji}^{-1}$ and we impose gauge invariance at each vertex $i$ of the graph, i.e.:
\begin{equation}
\psi_{\Gamma}(\{h_{ij}\})=\psi_{\Gamma}(\{g_ih_{ij}g_j^{-1}\})\qquad\forall g_i\in\mathbb G\;.
\end{equation}
Consider now a new Hilbert space given by
\begin{equation}
\mathcal H_{V}\cong L^2(\mathbb G^{d\times V}/\mathbb G^V)\;,
\end{equation}
whose generic element will be a function of $d\times V$ group elements
\begin{equation}
\varphi(\{g_i^a\})=\varphi(g_1^1,\dots,g_d^1,\dots,g_1^V,\dots,g_d^V)\;\in\;\mathcal H_V
\end{equation}
satisfying the gauge invariance at the vertices of the graph, i.e.: $\forall\alpha\in\mathbb G$,
\begin{equation}
\varphi(\dots,g_a^i,\dots,g_b^j,\dots)=\varphi(\dots,\alpha_ig_a^i,\dots,\alpha_jg_b^j,\dots)\;.
\end{equation}
As in LQG, the measure of the Hilbert space is taken to be the Haar measure. The interpretation of such functions is that each $\varphi$ is associated to a $d$-valent graph formed by $V$ disconnected components, each corresponding to a single $d$-valent vertex and $d$ 1-valent vertices, which are called \textit{open spin network vertices}.\\Given a closed $d$-valent graph $\Gamma$ with $V$ vertices specified by $L(\Gamma)$, a cylindrical function $\psi_\Gamma$ can be obtained by group averaging a wave function $\varphi$
\begin{align}\label{gluope}
\psi_\Gamma(\{h_{ij}^{ab}\})&=\int_\mathbb G\prod_{[(ia)(jb)]\in L(\Gamma)}d\alpha_{ij}^{ab}\;\varphi(\{g_i^a\alpha_{ij}^{ab};g_j^b\alpha_{ij}^{ab}\})\\ \nonumber
&=\psi_{\Gamma}(\{g_i^a(g_j^b)^{-1}\})
\end{align}
in such a way that each edge is associated with two group elements $g_i^a,g_j^b\in\mathbb G$. The integrals over $\alpha$ operate a ``gluing'' of the open spin network vertices corresponding to $\varphi$, pairwise along common links, thus forming the closed spin network represented by the closed graph $\Gamma$. Such a gluing can be interpreted as a symmetry requirement. Essentially, what we are saying is that we impose the function $\varphi$ to depend on the group elements $g_i^a,g_j^b$ only through the combination $g_i^a(g_j^b)^{-1}=h_{ij}^{ab}$ which is invariant under the group action, by the same group element, at the endpoint of two open edges to which these group elements are associated as showed in Fig. \ref{glu} for the simple example of the tetrahedral graph.
This shows that, only using functions $\varphi$, it is always possible to construct a generic function $\psi$ with all the right variables and symmetry properties, i.e., the space of functions $\psi$ is a subset of the space of functions $\varphi$.\\Moreover, using the Peter-Weyl decomposition theorem, we can give the corresponding formula in the spin representation which expresses the gluing of open spin network vertices and defines cylindrical functions for closed graphs as special cases of functions associated to a given number of them. Indeed, a cylindrical function $\psi_{\Gamma}$ can be decomposed as
\begin{align}\label{114}
%\begin{split}
&\psi_{\Gamma}(\{h_{ij}^{ab}\})=\\ \nonumber
&=\sum\psi_{\{m_{ij}^{ab}k_{ij}^{ab}\}}^{J_{ij}^{ab}}\prod_i\overline{C^{J_{ij}^{ab}\mathcal I_i}_{m_{ij}^{ab}}}C^{J_{ij}^{ab}\mathcal I_i}_{n_{ij}^{ab}}\prod_{[(ia)(jb)]}D^{(J_{ij}^{ab})}_{s_{ij}^{ab}n_{ij}^{ab}}\bigl(\{h_{ij}^{ab}\}\bigr)\\ \nonumber
&=\sum_{\{J\},\mathcal I}\tilde\psi^{\{J_{ij}^{ab}\},\mathcal I_i}\prod_iC^{J_{ij}^{ab}\mathcal I_i}_{n_{ij}^{ab}}\prod_{[(ia)(jb)]}D^{(J_{ij}^{ab})}_{s_{ij}^{ab}n_{ij}^{ab}}\bigl(\{h_{ij}^{ab}\}\bigr)
%\end{split}
\end{align}
where
\begin{itemize}
\item $J_{ij}^{ab}$ label the representations of the group $\mathbb G$ and $D^{(J)}$ are the corresponding representation matrices whose indices refer to the start and end vertex of the edge $[(ia)(jb)]$ to which the group element $h_{ij}^{ab}$ is attached;
\item $C^{\{J\}, \mathcal I}$ are the normalized intertwiners for the group $\mathbb G$, attached in pairs to the vertices, resulting from the gauge-invariace requirement, a basis of which is labelled by additional quantum numbers $\mathcal I$. These intertwiners contract all indices of both nodes and of the representation functions, leaving a gauge-invariant function of spin variables only. 
\end{itemize}
By using a similar decomposition for the function $\varphi$, the group averaging expression of $\psi$ in terms of $\varphi$ can be written as
\begin{widetext}
\begin{eqnarray} \label{115}
\psi_{\Gamma}(\{h_{ij}^{ab}&=&g_i^a(g_j^b)^{-1}\})=\int\prod_{[(ia)(jb)]}d\alpha_{ij}^{ab}\sum_{\substack{\{\vec J_i,\vec m_i\},\\ \mathcal I_i}}\varphi_{\vec m_i}^{\vec J_i,\mathcal I_i}\prod_i\biggl(\prod_{j\neq i}D^{(J_i^a)}_{m_i^an_i^a}(g_i^a\alpha_{ij}^{ab})\biggr)C_{\vec n_j}^{\vec J_j,\mathcal I_i}\\ \nonumber
&=&\sum_{\substack{\{J_{ij}^{ab}\},\\ \{m_i^j\},\\ \mathcal I_i}}\varphi_{\vec m_i}^{\vec J_i,\mathcal I_i}\prod_iC_{n_{ij}^{ab}}^{J_{ij}^{ab},\mathcal I_i}\prod_{[(ia)(jb)]}\delta_{J^a_i,J_j^b}\delta_{m^a_i,m_j^b}D^{(J_{ij}^{ab})}_{s_{ij}^{ab}n_{ij}^{ab}}(g_i^a(g_j^b)^{-1})
\end{eqnarray}
from which, comparing with (\ref{114}), we get the gluing formula in spin representation
\begin{equation}\label{gsr}
\psi^{\{J_{ij}\},\mathcal I_i}=\sum_{\{\vec m\}}\varphi_{\vec m_i}^{\vec J_i,\mathcal I_i}\prod_{[(ia)(jb)]}\delta_{J^a_i,J_j^b}\delta_{m^a_i,m_j^b}\;.
\end{equation}
\end{widetext}

This means that LQG states can be regarded as linear combinations of disconnected open spin network states with additional conditions enforcing the gluing and encoding the connectivity of the graph. Explicitly, Eq. (\ref{gsr}) shows that such conditions basically correspond to insert intertwiners given by the identity map at the bivalent vertices where the open links are pairwise glued.\\In order to deal with graphs with an arbitrary number of vertices, we consider the Hilbert space
\begin{equation}\label{new}
\mathcal H=\bigoplus_{V=0}^\infty\mathcal H_V\;.
\end{equation}
Eq. (\ref{gluope}), or equivalently (\ref{115}), shows that there is a correspondence between LQG states and states in $\mathcal H$. This is actually more than a correspondence at the level of sets of states since it is possible to prove that the scalar product in $\mathcal H_V$ for the special class of states corresponding to closed graphs induces the standard LQG kinematical scalar product for cylindrical functions $\psi_\Gamma\in\mathcal H_\Gamma$ based on a fixed graph (see \cite{LQG60} for details). This means that, assuming that the graph $\Gamma$ has $V$ vertices, $\mathcal H_\Gamma$ can be embedded into $\mathcal H_V$ faithfully, i.e., preserving the scalar product.\\ \\ Still, even though they agree exactly for each $\mathcal{H}_\Gamma$, it is important to stress the main differences between the new Hilbert space $\mathcal{H}$ and $\mathcal{H}^0_{kin}$.
\begin{itemize}
\item[\textbf{1)}] The Hilbert space $\mathcal H$ in (\ref{new}) is defined by taking the direct sum over all the Hilbert spaces $\mathcal H_V\supset\mathcal H_\Gamma$ with fixed number of vertices without introducing any cylindrical equivalence class. As such, unlike the LQG case, zero modes are now included in the Hilbert space.
\item[\textbf{2)}] In the new Hilbert space, states associated to different graphs are organized in a different way w.r.t. the LQG space. Indeed, states associated to graphs with different number of vertices are orthogonal, but those associated to different graphs but with the same number of vertices are not orthogonal.
\end{itemize}
The functions $\varphi(\vec g_1,\dots\vec g_V)$ can be understood as ``many-body'' wave functions for $V$ quanta corresponding to the $V$ open spin network vertices to which the function refers. Indeed, each state can be decomposed into products of ``single-particle''/``single-vertex'' states
\begin{equation}
\ket{\varphi}=\sum_{\{\vec\chi_i\}_{i=1,\dots,V}}\varphi^{\vec\chi_1\dots\vec\chi_V}\ket{\vec\chi_1}\otimes\dots\otimes\ket{\vec\chi_V}\;,
\end{equation}
which in the group representation reads as
\begin{equation}
\varphi(g)\equiv\braket{g|\varphi}=\sum_{\{\vec\chi_i\}}\varphi^{\vec\chi_1\dots\vec\chi_V}\braket{\vec g_1|\vec\chi_1}\dots\braket{\vec g_V|\vec\chi_V}
\end{equation}
where the complete basis of single-vertex wave functions is given by wave functions for individual spin network vertices, i.e. $\ket{\vec\chi}=\ket{\vec J,\vec m, \mathcal I}$, with
\begin{equation}
\psi_{\vec\chi}(\vec g)=\braket{\vec g|\vec\chi}=\biggl(\prod_{\ell=1}^dD^{(J_\ell)}_{m_\ell n_\ell}\biggr)C^{J_1\dots J_d,\mathcal I}_{n_1\dots n_d}\;.
\end{equation}
The normalization condition for the $\varphi$ is provided by
\begin{equation}
\int\prod_{v=1}^Vd\vec g_v\bar\varphi(\vec g_1,\dots\vec g_V)\varphi(\vec g_1,\dots\vec g_V)=\sum_{\{\chi_v\}}\bar\varphi^{\{\chi_v\}}\varphi^{\{\chi_v\}}\;,
\end{equation}
where we have used the normalization condition of single-particle wave functions
\begin{equation}
\int d\vec g\,\bar\psi_{\vec\chi\,'}(\vec g)\psi_{\vec\chi}(\vec g)=\delta_{\vec\chi\,',\vec\chi}\;.
\end{equation}
The functions $\varphi$ are exactly the many-body wave functions for point particles living on the group manifold $\mathbb G^d$, whose classical phase space is $(T^*\mathbb G)^d\cong(\mathbb G\times\mathcal G^*)^d$ which is also the classical phase space of a single polyhedron dual to a $d$-valent spin network vertex. The resulting picture of the microstructure of spacetime is thus based on glued pre-geometric fundamental building blocks. This is the general picture underlying the GFT formalism. This is even more evident in a 2nd quantized, Fock space reformulation fo the same Hilbert space $\mathcal{H}$, where building blocks are created and annihilated and their gluing corresponds to interactions of combinatorial nature \cite{LQG60,LQG27}. Along with  the examples given in Section \ref{2wl}, also the gluing of open spin networks can be naturally understood in terms of entanglement of their spin network degrees of freedom, so that the GFT Hilbert space fits very well our pre-geometric approach to quantum spacetime and the idea of reconstructing geometry from entanglement.

\end{document}